\pdfoutput=1
\documentclass[twocolumn]{scrartcl}

\usepackage[utf8]{inputenc}
\usepackage[T1]{fontenc}
\usepackage[english]{babel}
\usepackage{graphicx}
\usepackage{amssymb}
\usepackage{amsmath}
\usepackage{subfig}
\usepackage{color}
\usepackage{textcomp}
\usepackage{url}
\usepackage{setspace}
\usepackage[normalem]{ulem}
\usepackage{authblk}

\newenvironment{keywords}%
   {\begin{trivlist}\item[]{\bfseries\sffamily Keywords:}\ }
   {\end{trivlist}}

\begin{document}

\title{Local Fiber Orientation from X-ray Region-of-Interest Computed
  Tomography of large Fiber Reinforced Composite Components}

\author[a]{Thomas Baranowski}
\author[b,c]{Dascha Dobrovolskij}
\author[d]{Kilian Dremel}
\author[d]{Astrid H{\"o}lzing}
\author[e]{G{\"u}nter Lohfink}
\author[c]{Katja Schladitz\thanks{Corresponding author:\\ \texttt{katja.schladitz@itwm.fraunhofer.de}}}
\author[d]{Simon Zabler}

\affil[a]{Ford Research and Innovation Center Aachen, Ford-Werke GmbH}
\affil[b]{Hochschule Darmstadt, Schöfferstraße 3, 64295 Darmstadt, Germany}
\affil[c]{Fraunhofer-Institut für Techno-und Wirtschaftsmathematik, Fraunhofer-Platz 1, 67663 Kaiserslautern, Germany}
\affil[d]{Fraunhofer-Entwicklungszentrum Röntgentechnik, Josef-Martin-Weg 63, 97074 Würzburg, Germany}
\affil[e]{Montaplast GmbH, Morsbach, Germany}

\maketitle


\begin{abstract}
  The local fiber orientation is a micro-structural
  feature crucial for the mechanical properties of parts made from
  fiber reinforced polymers. It can be determined from micro-computed
  tomography data and subsequent quantitative analysis of the
  resulting 3D images. However, although being by nature
  non-destructive, this method so far has required to cut samples of a
  few millimeter edge length in order to achieve the high lateral
  resolution needed for the analysis.
   
  Here, we report on the successful combination of region-of-interest
  scanning with structure texture orientation analysis rendering the
  above described approach truly non-destructive. Several regions of
  interest in a large bearing part from the automotive industry made
  of fiber reinforced polymer are scanned and analyzed. Differences of
  these regions with respect to local fiber orientation are
  quantified. Moreover, consistency of the analysis based on scans at
  varying lateral resolutions is proved.  Finally, measured and
  numerically simulated orientation tensors are compared for one of
  the regions.
\end{abstract}

\section{Introduction}
Many structural components in the automotive and aircraft industries
are made from fiber reinforced plastic (FRP) composite material.  The
fibers can be made of glass, carbon, or other materials, they can be
short cut, long or even continuous, of various thicknesses and volume
concentrations.  Typically, the fiber component comprises 10 to 35\,\%
of the volume for injection molded materials (e.~g. in
\cite{andrae-wirjadi2014}) and up to 50-60\% for laminar FRP
(e.~g. \cite{requena:2009,sencu:2016,heieck2017,emerson:2017}) and
consists of 6 to 15\,\textmu m thick carbon fiber bundles or 10 to
18\,\textmu m thick glass fibers (e.~g. 10\,\textmu m in
\cite{andrae-wirjadi2014}, 12\,\textmu m in \cite{musiko2019},
18\,\textmu m in \cite{hannesschlaeger2015}).
  
When the components are molded, usually thermoplastics like
polypropylene (e.~g. \cite{wirjadi2014:multilayer}), polybutylene
terephthalate (e.~g. \cite{musiko2019}), polyamide 66
(e.~g. \cite{ayadi2016}), polyamide 6, acrylonitrile butadiene styrene
are used as host material to which the fibers are added. Since FRP
components generally undergo mechanical and/or thermal stresses during
their service life, their load bearing capacity/strength is of
critical importance for the components' design.

The structural properties of injection molded FRP
materials are locally anisotropic due to the fiber component being
anisotropically oriented \cite{advani87,miled-numsim2008},
\cite{ayadi2016} (for short fibers), \cite{fliegener2017} (for long
fibers). This anisotropy in turn is caused by the fibers moving with
the liquid flow in the mold \cite{bernasconi2012analysis}. For complex
shaped parts, the resulting fiber orientations are difficult to
predict and control. As a consequence, structurally weak spots 
or areas can appear and may lead to early failure of the component. In
order to avoid this, components are often designed too thick.  This in
turn thwarts the weight saving intention in using FRP.

Numerical simulations can predict the liquid flow \cite{corheos2013}
and thus indicate critical areas where material weaknesses might
occur. Yet, these simulations are not perfect and need
validation. Moreover, prediction of the local materials properties
relies on orientation information as input, typically in the form of
the 2nd order orientation tensor \cite{advani87}.

Fiber orientations can be analyzed essentially by four types of
methods. Historically, before X-ray micro computed tomography (\textmu
CT) became widely available, fiber orientations were evaluated through
image analysis of polished 2D sections. See \cite[Section
11.6.4]{KMS95} for a summary of stereological methods based on
counting intersections in slices at varying angles and
\cite{mlekusch1999,eberhardt2001} for orientations deduced from the
shape of the observed elliptical cross-sections. These methods are not
only destructive but the latter suffers also from the need for rather
high resolutions and ambiguities due to the fact that there are always
two 3D orientations generating the same cross-sectional
ellipse. Therefore, critical areas of damaged/failed thermoplastic
parts (or field returns) are nowadays predominantly analyzed by CT
techniques, nevertheless for detailed verification optical microscopy
of polished micro-sections is still commonly used.

X-ray micro computed tomography (\textmu CT) can be employed to
analyze non-destructively the fiber orientation in structural
components. Based on the resulting three-dimensional images, fiber
orientations can be analyzed for the whole field of view (FoV) by the
mean intercept length (MIL, \cite{bernasconi2012analysis}) or by
measuring the length of generalized projections and obtaining the
orientation distribution via the inverse cosine transform
\cite[Section 5.4]{ohser-schladitz09book}. Both methods are designed
to be applied to the whole FoV, they can however be localized by
applying them to small sub-volumes. Nevertheless, in order to get an
orientation in each voxel, one would have to center the sub-volume in
each voxel. While inverting the cosine transform faces numerical
instabilities, the MIL method has been applied successfully, see
\cite{bernasconi2012analysis} also for a comparison to analysis of 2D
virtual slices from the CT data as well as of 2D images of polished
surfaces.

Local fiber orientation can of course be determined via single fiber
segmentation. These approaches are typically very demanding with
respect to image quality (contrast as well as lateral resolution),
fiber volume fraction and spatial arrangement of the fibers. On the
other hand, if successful they yield additional valuable information,
in particular the fiber length distribution \cite{salaberger2011,
  tessmann:2010} or positions with respect to failure regions
\cite{kronenberger2018}. Single fiber segmentation usually relies on
tracking fiber center lines or cross-sections from slice to slice
\cite{requena:2009,emerson:2017} or local approximation of fibers by
line segments \cite{eberhardt2002automated,sandau07}, ellipsoids
\cite{altendorf09:_fiber_separ}, or cylinders
\cite{shen2004,mishurova2018,pinter:2016}. \cite{altendorf09:_fiber_separ}
reconnects fragments based on local orientation, while
\cite{shen2004,mishurova2018} just use the fragment orientations. All
these approaches have in common that they demand rather slow
orientation changes within one fiber (see the detailed discussion in
\cite{kronenberger2018}) and the fiber diameter to be resolved by at
least 8 voxels.  Vigui{\'e} \cite{viguie2013} and Kronenberger
\cite{kronenberger2018} do not need the former but \cite{viguie2013}
thrives on high image quality as provided by tomography using
synchrotron radiation and both rely on resolutions of 8 voxels per
fiber diameter and more. Pinter \cite{pinter:2016} claims 5 voxels per
diameter to be sufficient, whereas the efficiency of the circular
voting filter drops significantly for the lower resolution of 3 voxels
per fiber diameter. In \cite{ayadi2016}, orientations of short fibers
or clusters of them are used to derive the homogenized orthotropic
behaviour for cuboidal sub-volumes for use in FEM simulations.

Here, we concentrate on estimation of the local fiber orientation in
the sense of assigning an orientation vector to each voxel belonging
to the fiber system without fiber separation. For this purpose,
several methods based on local approximation of fibers by ellipsoids
\cite{robb07,altendorf09,wirjadi09a} and on local gray value
derivatives of first \cite{krause10} and second order \cite{wirjadi16}
have been proposed. The first order gray value derivatives are
subsumed into the so called structure tensor and are applied e.~g. by
\cite{denos2018} and in VG STUDIO MAX, see
e.~g. \cite{belmonte2016}. Note that the structure tensor is not the
orientation tensor as described by \cite{advani87} and given in
Equation~\eqref{eq:orientation-tensor} below. The second order
derivatives form the so called Hessian matrix and are applied e.~g. in
MAVI, see e.~g.  \cite{wirjadi2014:multilayer}.  The rationale behind
both methods is that locally the fiber orientation is the one in which
gray values change or are curved the least. In \cite{wirjadi16}, all
four methods are compared comprehensively based on simulated single
fibers with a diameter of 10 voxels. In applications, the derivative
based methods usually use 2-4 voxels per diameter
\cite{hannesschlaeger2015,wirjadi2014:multilayer,wirjadi16,
  tessmann:2010} while \cite{altendorf09} rather demands 10. An
exception is \cite{denos2018} applying the structure tensor at fiber
bundle instead of fiber level thus allowing for voxel sizes of
50\,\textmu m. In \cite{prade2017}, the authors present a different
technique, the dark field scanning, for derivation of fiber
orientation results in FRP. This method allows for even coarser
resolutions of 86\,\textmu m.
\begin{figure*}[ht!]
 \includegraphics[width=\textwidth]{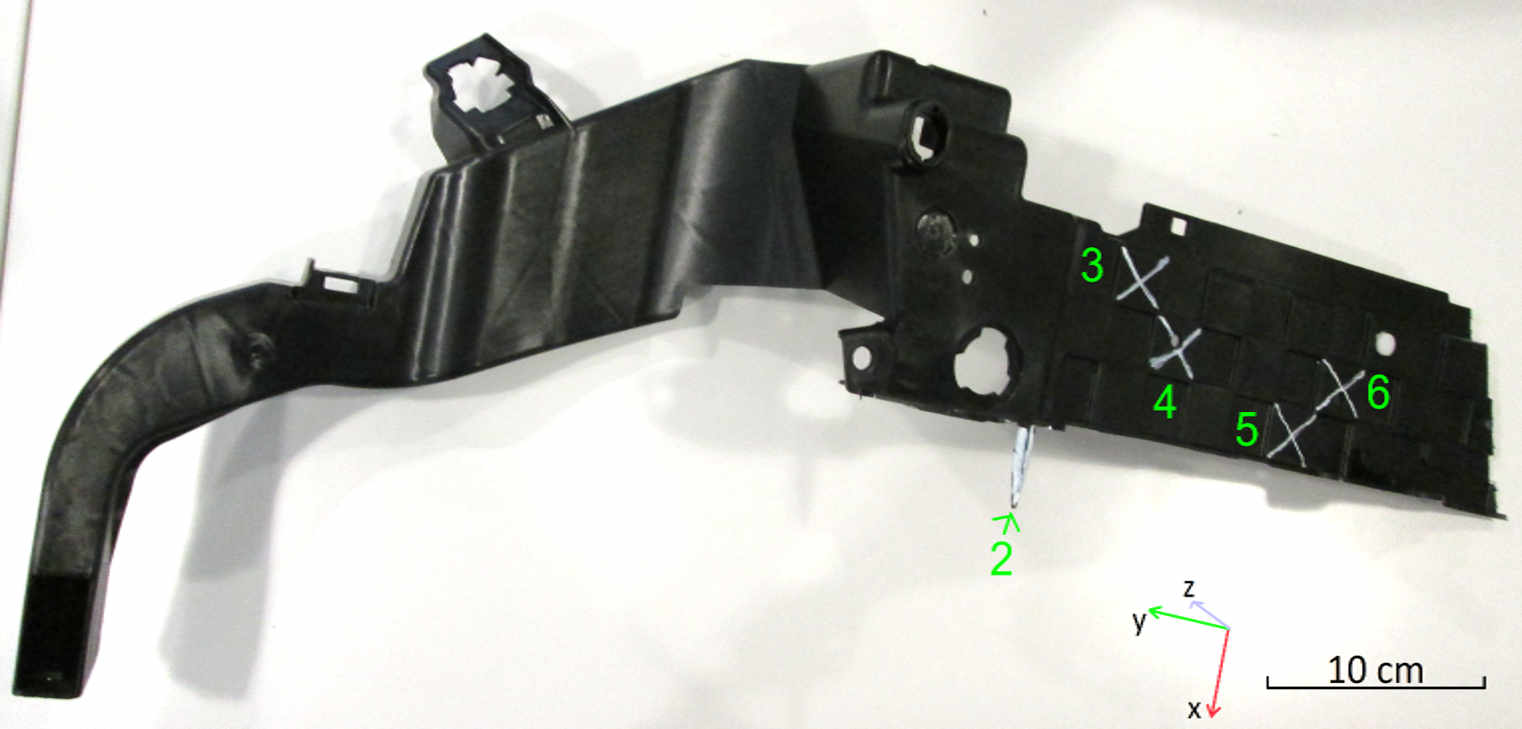}
 \caption{Bearing component made of polypropylene with 30\,wt\%
   reinforcing long glass fibers. Regions of interest (RoI) are marked
   white. As indicated, the analyzed component is mostly oriented in
   the x-y-plane and positive z is the thickness
   direction. Throughout, this coordinate system is used.}
\label{fig:ROI_Ford_Bauteil}
\end{figure*}

All local and single fiber based orientation analysis methods
described above except \cite{denos2018, prade2017} require the fiber
diameter to be sampled by at least 2-4 voxels
\cite{redenbach12:_beyon_imagin,wirjadi2014:multilayer,PINTER201826}.
Sampling the fibers coarser than this causes them to crumble in the
digital image. That is, a fiber sampled at less than 2 voxels for its
diameter is endangered to form more than one connected
component. Thus, similar to the microscopic imaging of planar sections
- \textmu CT is limited in its FoV to some mm$^3$. E.~g., if glass
fibers of diameter 10\,\textmu m are analyzed, orientation analysis
limits the voxel size to at most 5 \,\textmu m and consequently the
FoV to 10\,\textmu m/3$\times$2\,048$\approx$7\,mm size.  This FoV
usually covers the component's thickness. However, the lateral
dimensions of injection molded parts can be as large as some
meters. So far, we used CT to analyze fiber orientations in small
molded parts. These results have been used to verify simulation
results of injection molding simulation software. Large molded parts
could not be analyzed due to the limitations of CT devices used in the
polymer industry.  The part analyzed in this paper, shown in
Figure~\ref{fig:ROI_Ford_Bauteil}, is about 1\,m long.

As a consequence, very often the sample size is reduced to match the
FoV determined by the needed resolution by cutting
small pieces of a few millimeters edge length from
the part thereby rendering \textmu CT an utterly destructive
inspection technique \cite{wonisch2014,ayadi2016}.

With this work we demonstrate two strategies which aim at
employing \textmu CT for fully non-destructive analysis of local fiber
orientation in injection molding FRP parts:
\begin{enumerate}
\item The imaging technique is applied as region of interest CT (RoI
  CT) \cite{garcea2017}. 
  That is, the components are not cut and the FoV can be
  significantly smaller than the components size, at least in one
  dimension. 
\item Instead of the local fiber orientation distribution, a local
  ``texture orientation'' is derived. More precisely, the local
  orientation is determined in small cubic sub-volumes (boxes). Boxes
  not containing enough voxels belonging to the fiber component are
  not taken into account. Thus, the spatial sampling of the fiber
diameter can be reduced to less than one voxel, allowing for an FoV 10
times larger than required for a local fiber orientation analysis.
\end{enumerate}
By combining these two aspects into one novel method the analysis of
local fiber orientation in meter-sized glass FRP components becomes readily
feasible.

While in synchrotron \textmu CT experiments RoI (axial or
laminography) scans are performed routinely on carbon FRP material
\cite{bull2013} this acquisition mode is still new in laboratory
\textmu CT. One reason for this is the compactness of commercial
\textmu CT scanners which, given a high geometric magnification, leave
little to no space between X-ray source window and sample. The present
scanner uses a variable source-detector distance which allows for RoI
CT in samples of 100 mm width while maintaining a high magnification
and microscopic voxel sampling.
\begin{figure*}[!ht]
\begin{center}
\includegraphics[width=0.9\textwidth]{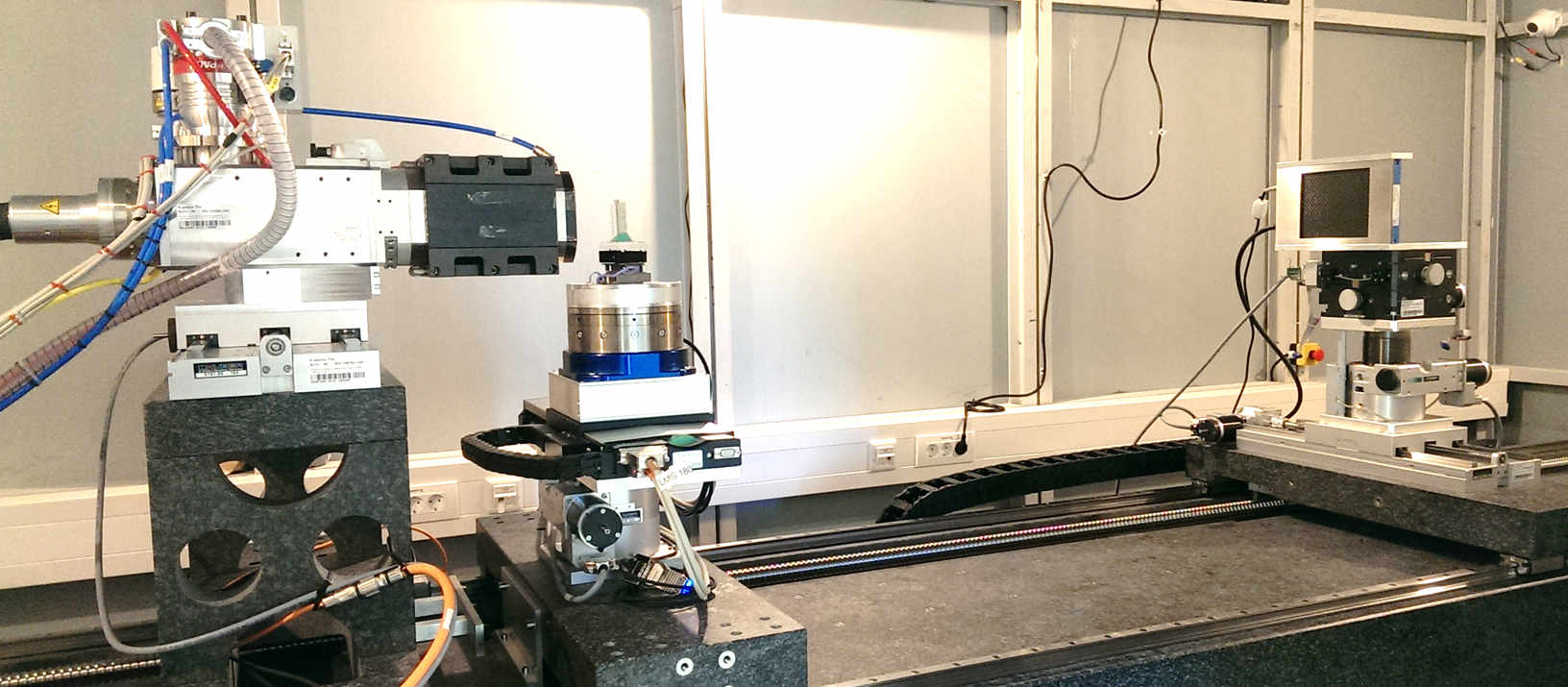} 
 \end{center}
 \caption{RoI CT scanner MetRIC at EZRT in Würzburg.}
\label{fig:Anlage_Wurzburg}
\end{figure*}

\section{Materials and methods}
The object used to demonstrate our technique is a long glass fiber
composite carrier, see Figure~\ref{fig:ROI_Ford_Bauteil}. Built into
the upper front end of the car, this part fulfills several functions:
It carries the hood damper and yields mounting points for the lights
as well as for the radiator package. The latter is provided with air
by the carrier, too. Moreover, jointly with the crash absorber, the
carrier contributes to fulfilling the legal regulations
w.r.t. pedestrian protection.

In a polypropylene matrix, 30\,wt\% glass fibers of 10-20\,\textmu m
thickness and 10-15\,mm length before processing are embedded.  The
carrier has been scanned several times with different
parameters. Thus, data has been acquired at several resolutions
revealing different local features of the particular regions.  The
scanned regions were chosen to evaluate the effect of flow on fiber
orientations along the part.  Table~\ref{tab:regions} summarizes
physical and digital sizes of the scans.  The largest analyzed RoI
scan A3.1 was scanned at the coarse resolution of 44\,\textmu m/voxel
edge by the Tomosynthesis scanner at a the Fraunhofer Institute in
Fürth (EZRT), Germany.  This scanner has a maximum source-detector
distance of 2 m, while its detector has 100 \textmu m pixel pitch.
Thus the scanner does not allow for very fine voxel
samplings ($<$ 5\textmu m) while maintaining a large source-object
distance (in the present case at least 100 mm). We therefore used the
metRIC scanner at EZRT Würzburg for the remaining scans. The metRIC
allows for voxel samplings down to 2.33 \textmu m with 100 mm
source-object distance thanks to its very large detector X-axis (up to
3.3 m) and a pixel pitch of 74.8 \textmu m. Thus, metRIC allows for
higher resolutions while coping with large size of the scanned
component. So that, we scanned the RoI A2, A4 and A5 at higher
resolutions of 10-20\,\textmu m/voxel edge. In order to compare the
local analysis at several resolutions, the region A3.2, has been
imaged at coarse (45\,\textmu m/voxel, A3.2m), intermediate
(21\,\textmu m/voxel, A3.2h), and high resolutions (10\,\textmu
m/voxel, A3.2uh). The scans A3.2m, A3.2h and A3.2uh have been acquired
by changing exclusively the source-object-distance/source-detector-
distance leading to the corresponding dimensions of the scanned RoI.
Finally, RoI A3.3 has been scanned at the highest resolution of
3\,\textmu m/voxel edge.  Next, we describe the CT-data acquisition
set-ups. Afterwards, we present the orientation analysis results for
the CT-data and compare them with
Moldflow\textsuperscript{\textregistered} simulations.
\begin{table*}[ht!]
\centering
  \begin{tabular}{|c|c|c|r|r|}
  \hline
  RoI & CT device & dimensions in & voxels & voxel size\\
        &  & x-y-plane [cm$^2$]   &        & [\textmu m]\\
  \hline\hline 
   A2 & MetRIC & $2.7\times0.9$ & $\phantom{1\,}604\times\phantom{1\,}200\times\phantom{1\,}205$ & $45$\\ \hline
   A3.1 & Tomosyn & $6.2\times4.8$ & $1\,422\times1\,102\times\phantom{1\,}450$ & $44$\\ 
   A3.2m & MetRIC & $4.9\times2.5$ & $1\,084\times\phantom{1\,}556\times\phantom{1\,}172$ & $45$\\
   A3.2h & MetRIC & $2.0\times1.4$ & $\phantom{1\,}949\times \phantom{1\,}686\times \phantom{1\,}307$ & $21$\\ 
   A3.2uh & MetRIC & $1.0\times0.7$ & $\phantom{1\,}972\times \phantom{1\,}726\times \phantom{1\,}543$ & $10$\\ 
   A3.3 & MetRIC & $0.6\times2.3$ & $1\,944\times 8\,832\times 1\,700$ & $3$\\ \hline
   A4 & MetRIC & $3.0\times1.9$ & $1\,860\times1\,150\times\phantom{1\,}360$ & $17$\\ \hline
   A5 & MetRIC & $5.2\times5.6$ & $1\,210\times1\,300\times\phantom{1\,}250$ & $44$\\ 
  \hline
 \end{tabular}
\caption{RoI scanned at varying resolutions.}
\label{tab:regions}
\end{table*}

\subsection{X-ray approach for entire components (RoI CT)}
RoI CT of large glass FRP components requires certain degrees of freedom in
the CT scanner as well as sufficient space for displacement of the
sample. This is realized in the Tomosynthesis scanner. The
scanner allows for precise x, y, and z movements over more than one
meter range, thereby placing FoV at any position and of variable size
and detail between X-ray source and detector.  The X-ray source is an
open microfocal transmission anode which provides X-ray spot sizes
down to 1\,\textmu m. The X-ray projections of the sample are recorded
on a digital detector array which covers an area of 40\,cm$\times$40\,cm
(Varian PaxScan). The sample is mounted vertically on the object table
and the RoI is positioned on the marked positions covering 5\,cm FoV
which are sampled at 22.4\,\textmu m/voxel (geometric magnification
4.45$\times$). Binning of two detector pixels results in the final effective
voxel edge length of 45\,\textmu m.  Reconstruction of the volume
images from 3\,000 projections is achieved through standard Feldkamp
back-projection. 

\begin{figure}[!ht]
\begin{center}
\includegraphics[width=\columnwidth]{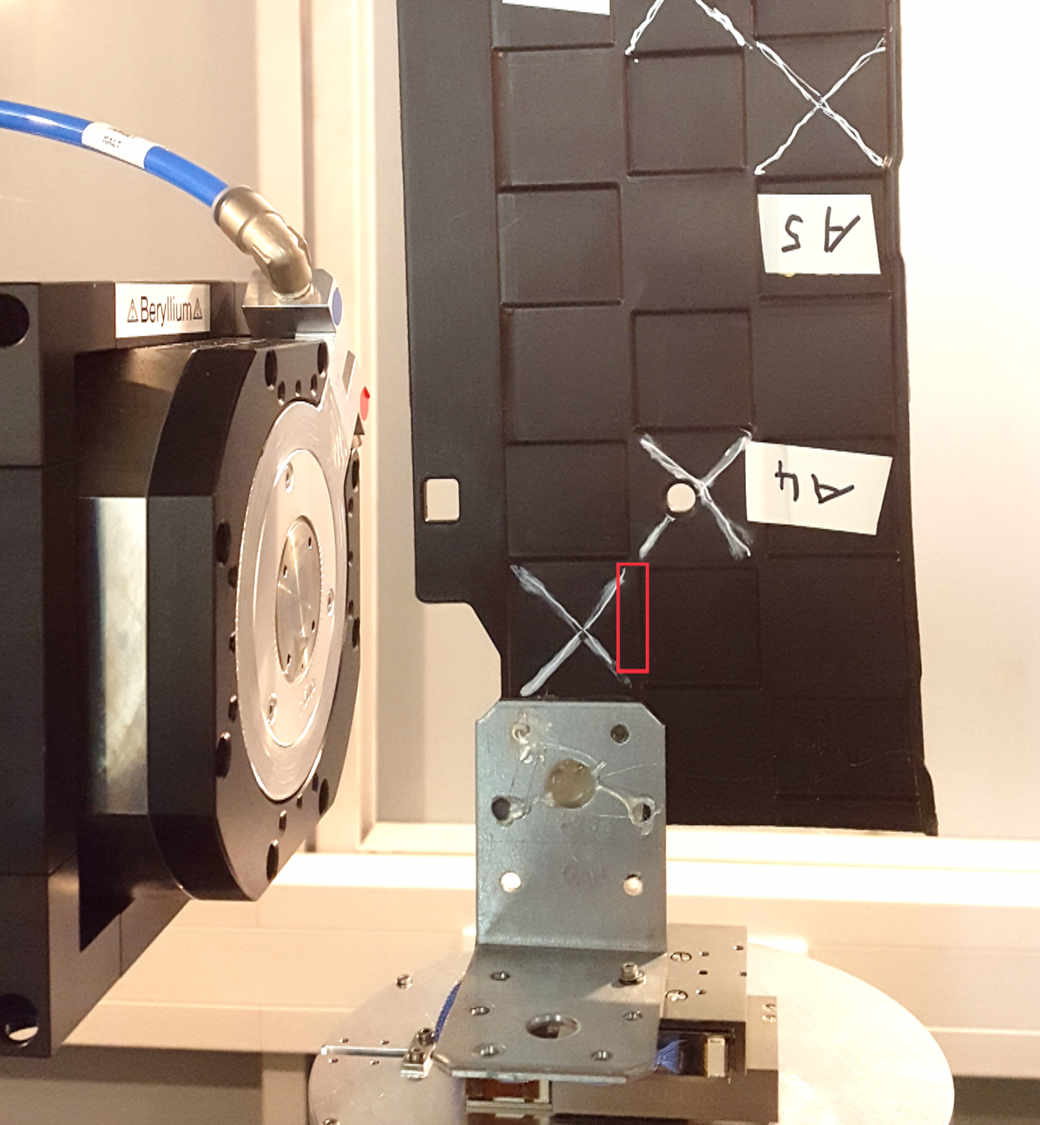}
 \end{center}
 \caption{The red box marks the RoI which has been scanned at finest
   resolution, divided into six sub-volumes. The stack of six scans
   covers approximately the entire red box.  Figure
   \ref{fig:vr-slices-ct-1}(a) shows the volume rendering of the
   segmented fiber component of the whole RoI A3.3.}
\label{fig:Anlage_Wurzburg_highres}
\end{figure}

In addition to the low and medium resolution scans of RoI A2, A3.1,
A3.2, A4 and A5, we applied high resolution local tomography to the
same sample. The RoI A3.3 covers the region marked by the elongated
red box in Figure \ref{fig:Anlage_Wurzburg_highres}.  The RoI CT scanner
MetRIC (see Figure \ref{fig:Anlage_Wurzburg}) has been designed and
constructed recently at the EZRT in Würzburg, Germany. The X-ray
source is an X-RAY WorX microfocus transmission anode (XWT-190-THCE
PLUS) which can be operated at up to 190 kV acceleration voltage and
provides a spot size down to below 1\,\textmu m with the
high-resolution target (1\,\textmu m W on 250\,\textmu m Be). The
setup comprising source, sample manipulator, and detector offers 10
degrees of freedom and a precisely encoded positioning of the sample
and probed RoI. The flat-panel detector (Perkin Elmer Dexela 1512 NDT,
14-bit CMOS, Gd$_2$0$_2$S:Tb DRZS-scintillation screen) features a
pixel size of 74.8\,\textmu m on a sensitive area of
154.4\,mm$\times$114.9\,mm. Especially the extended movement of up to
2.5\,m of the detector stack along the horizontally oriented X-ray
direction as well as of the X-ray source (up to 1.3\,m) enhances an
optimized arrangement of encoded source, sample, and detector
positions for well-balanced magnification
(source-object-distance/source-detector-distance), resolution and
scanning time of each sample and RoI. The encoding of all axes enables
CT scans without manual movement of the sample and the RoI can be
reproduced at any time. The encoded x-y-stage on top of the rotary
table as well as the z-movement of the sample stack allow for
selection of several RoI without repositioning the
sample. Consequently, all data is generated automatically in the same
coordinate system.

Our scanner allows for an extensive focus-detector distance (here
3.3\,m) which in turn enables voxel samplings as small as 3\,\textmu
m/voxel, even in parts which are several cm wide. Moreover, on top of
the air bearing rotation, a piezo positioning system lets the user
choose an arbitrary RoI with sub-\textmu m precision.

The minimum resolution of all CT scanners depends on the sample size.
Tomosynthesis as well as MetRIC enable resolutions down to
2-3\,\textmu m.  Tomosynthesis offers additionally the option to scan
very large samples with high resolution. On the other hand, MetRIC has
a smaller focal spot size reaching even higher resolutions for smaller
samples.  The degree of freedom for movements and total distances is
higher and thus allows more variable sample geometries.

Although the 3D volume reconstruction of RoI CT data gives the chance
to achieve high resolution images, it suffers from a few specific
problems. If an area of the scanned object overlaps the
reconstructable FoV in direction of the beam at a certain
angle, this area will become part of the corresponding
radiography. Therefore, in the image reconstruction process, this outer
area will become part of the reconstructed volume. Moreover, problems
occur when the projection of the scanned object is bigger than the
horizontal detector size. Filtered back-projection of the
radiographies involves high-pass filtering. As a consequence, the
detector edge leads to reconstruction artifacts near the boundaries of
the imaged volume.

Both problems are treated adequately at MetRIC. The
scan is performed “on-the-fly” using a non-stop rotational
movement. That way, parts of the sample that are located outside the
FoV will move faster during object rotation. Hence, 
the blurring of these areas is increased proportional to
their distance to the center of rotation. In the resulting
image slices, this leads to a constant gray value offset only.
The horizontal overlap is handled by padding as the high-pass 
filter used for backprojection is less sensitive of the projected 
image. Thus, cupping artifacts in the outer areas of the
reconstructed slices are avoided.

\subsection{Measuring local fiber orientation from 3D image data}
\label{sec:locFiberOrientation}
In this paragraph, we shortly summarize the method
  for local 3D fiber orientation analysis based on 2nd order gray
  value derivatives. That is, the method based on an eigenvalue analysis of the
  Hessian matrix in each voxel of a 3D image.

State of the art methods for analyzing the fiber
  orientation in \textmu CT images of FRP parts determine the fiber
  orientation in each voxel without need to identify individual fibers
  \cite{krause10,PINTER201826,wirjadi16,riesz-paper2019}. More precisely, in each voxel
  belonging to the fiber system, a local fiber orientation is
  derived. These voxel-wise measurements
  yield the volume weighted orientation distribution of the fiber system
  observed in the 3D image. As discussed in the Introduction, in
  general, 
  segmenting individual fibers requires higher resolutions than the
  voxel-wise orientation analysis. The latter being reported to work
  at spatial sampling of the fiber diameter by 2-3 voxels
  (\cite{redenbach12:_beyon_imagin,wirjadi2014:multilayer,PINTER201826}), allows for an FoV
  considerably larger than required for single fiber analysis.

Here however, in the coarser scans, the mean fiber diameter of
approximately 10\,\textmu m is sampled by less than one voxel. Due to
the local orientations not rapidly changing
  spatially, local orientation analysis is nevertheless possible, see
  \cite{cost-wisa-paper-2016} analyzing bundles of glass fibers in
  sheet molding compound samples at a nominal resolution of
  17.3\,\textmu m in virtual 2D slices using the method from
  \cite{robb07} and \cite{denos2018} applying the structure tensor
  \cite{krause10} to prepreg platelet compression molded samples for
  3D orientation analysis at 50\,\textmu m nominal
  resolution. The gray value of an image voxel represents in
  that case the averaged energy absorbed by several neighboring
  fibers.

\subsection{Local fiber orientation analysis}
The regions scanned are not simply cuboidal. Thus, first of all, masks
for the RoI of the part are derived from the CT image data. To this
end, solid matter is separated from the surrounding air by a manually
chosen global gray value threshold. The resulting rough edges are
smoothed by a morphological opening with a 3$\times$3$\times$3 voxel
cube.

The local orientations are measured in each voxel
exploiting the second order partial derivatives of the local gray
values. That is, the Hessian matrix is computed in each
voxel. Subsequent analysis of the eigenvalues of the
  Hessian yields the local orientation of bright locally fiber like
  structures as the eigenvector corresponding to the smallest (in
  magnitude) eigenvalue. Following \cite{wirjadi16}, we define a 
  fiber like structure to be a subset of a dilated random fiber system
  which in turn is a collection of rectifiable curves. See
  \cite{mecke80,nagel83} for mathematical background. The idea behind
  the eigenvalue analysis is that standing on a (gray value) mountain
  ridge, the orientation of the ridge is the one in which the (gray
  value) relief is curved the least \cite{eberly94}. In
  \cite{wirjadi16}, this method has been proven to be equivalent to
  the structure tensor based one of \cite{krause10} and to outperform
  methods discretizing the orientation space, namely orientation
  derived from maximal response of anisotropic Gaussian filters
  \cite{robb07} or from the moments of intertia \cite{altendorf09}.

Here, the Hessian matrix based method is slightly altered. In
\cite{wirjadi16}, the fiber diameter is assumed to be known.
Calculation of the 2nd order partial gray value
  derivatives in each voxel is proceeded by smoothing with a Gaussian
  filter whose parameter is chosen to meet exactly the fiber
  radius. This choice is motivated by the interpretation of a bright
  glass fiber within a darker matrix forming a ridge in the gray value
  relief and the desire to observe the highest points of the ridge
  exactly at the center line of the fiber. Recent experiments
  \cite{riesz-paper2019} shed some doubt on this empirically deduced
  rule of thumb and this issue is currently being investigated.
  Clearly, choosing the width of the Gaussian as the fiber diameter is
  impossible if the diameter is resolved with less than 3
  voxels. Thus, in these cases, a minimal smoothing filter with a
  3$\times$3$\times$3 voxel mask approximating a Gaussian is used.

The presented local orientation analysis is based on the mathematical
concept of the typical point of a random closed set
\cite{ohser-schladitz09book,schn:wei08,KMS95}. Very roughly speaking,
one looks at the world from a point chosen "uniformly" within the
random set. As long as the fibers are of equal thickness and do not
intersect, the resulting distribution of the fiber orientation in this
typical point is the same as if just the one-dimensional fiber cores
are taken into account. Let $R$ be the distribution of the local fiber
orientation in the typical point. That is, $R$ is a probability
measure on the space of direction -- the upper half-sphere $S^2_+$. 
The 2nd order orientation tensor \cite{advani87} can be
interpreted as the 2nd moment of $R$. Let $u_i,\quad i\in\{x,y,z\}$
denote the component of some normalized direction vector $u$ in
coordinate direction $i$. Then the second order orientation tensor is
defined as $\left( a_{ij} \right)$ with
\begin{equation} 
a_{ij}=\int_{S_+^2} u_i u_j~R\left(du\right), \quad i,j\in\{x,y,z\},
\label{eq:orientation-tensor}
\end{equation}
see \cite{wirjadi16}.

Finally, the local orientation information is exploited for
those voxels assigned to the fiber system by a global gray value
threshold, only. This threshold is found by multiplying Otsu's
threshold \cite{otsu79} by 1.25, the rationale behind that being that
the observed orientation distribution is not distorted if voxels at
the fiber edges are systematically not taken into account.

An eigenvalue analysis for the 2nd order orientation tensor yields the
preferred local direction as the eigenvector to the largest eigenvalue
$\ell_{{\mathrm max}}$ as well as an index reflecting the strength of
anisotropy, see \cite{cost-wisa-paper-2016}. More precisely, consider
$\alpha = 1-\ell_{{\mathrm min}}/\ell_{{\mathrm max}}$, where
$\ell_{{\mathrm min}}$ is the smallest eigenvalue of the orientation
tensor. This index assumes values in the range $[0,1]$ with $0$
indicating perfect isotropy. A value of $1$ is achieved in two cases -
perfect unidirectional fibers or a transversally isotropic fiber
system. Tensile tests for glass FRP reported in
\cite{cost-wisa-paper-2016} suggest that samples with $\alpha < 0.6$
behave as isotropic samples. Thus calculating a mean fiber direction
is not sensible in this case.

\section{Results}
\label{sec:results}
Here, we present first the results of the RoI CT scans by volume
renderings and slice views in Section~\ref{sec:roi-ct-scans}. The
following Section~\ref{sec:local-fiber-orient} contains the 3D
orientation analysis results, both locally as well as averaged for
each region. Finally, for regions A3.2 and A3.3, local fiber
orientations deduced from the image data are compared for several
resolutions in Section~\ref{sec:comp-resolution} and for A3.3 to those
obtained by numerical simulation of the injection molding process, see
Section~\ref{sec:comp-local-fiber}.
\begin{figure*}[ht!]
 \subfloat[A3.3, volume rendering]{\includegraphics[width=\columnwidth,trim=0mm 0mm 0mm 25mm]{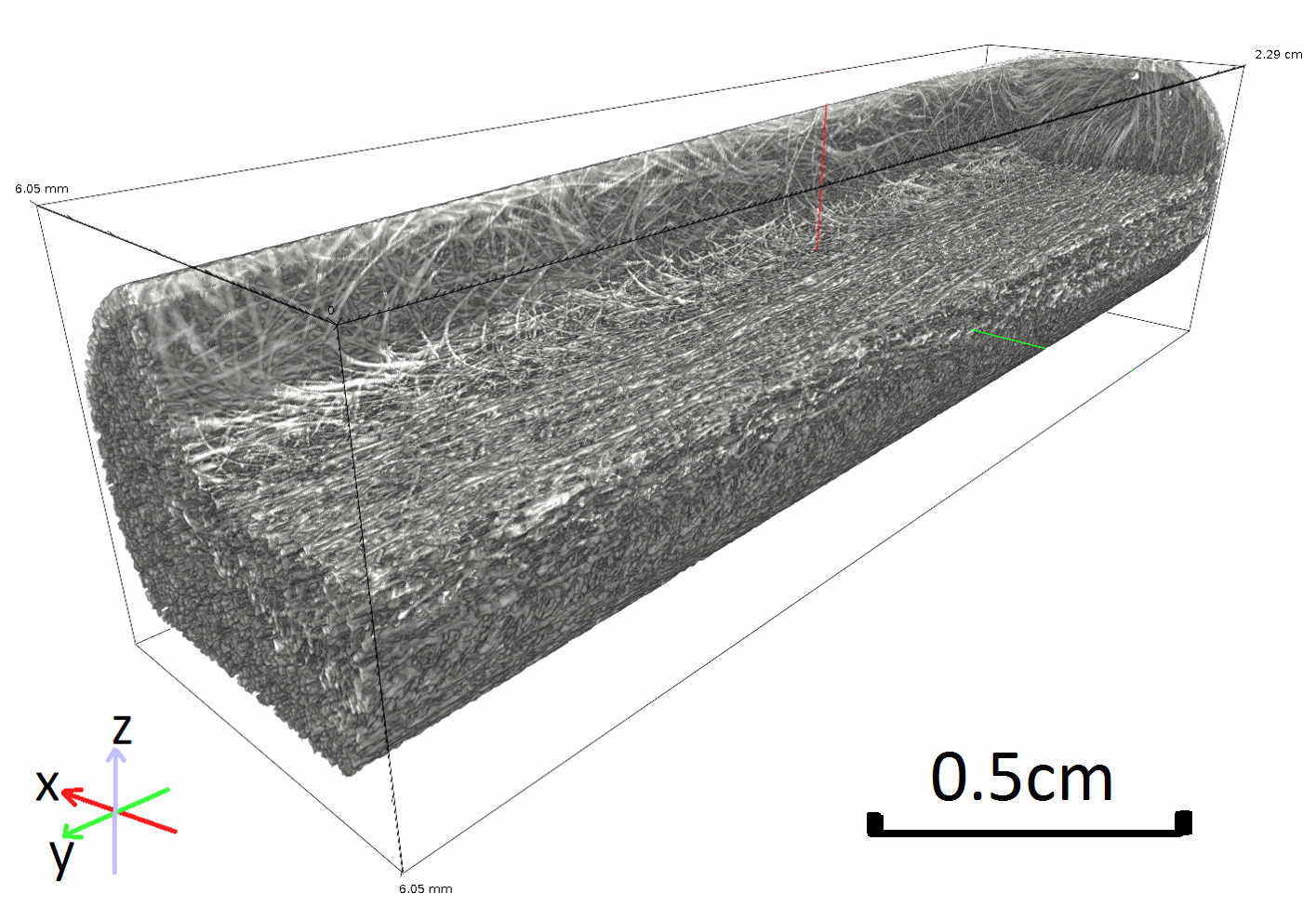}}\hfill
 \subfloat[A3.3, x-y-slice]{\includegraphics[width=\columnwidth]{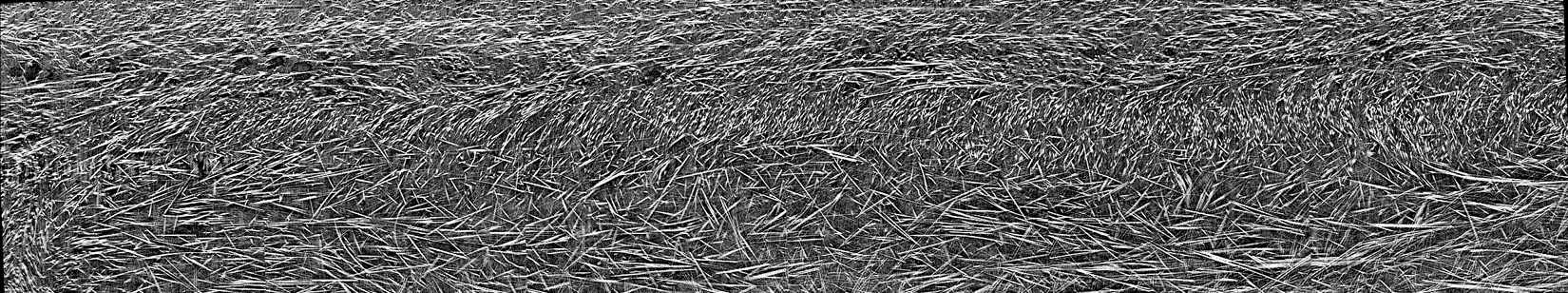}}\\
 \subfloat[A3.2uh, volume rendering]{\includegraphics[width=\columnwidth]{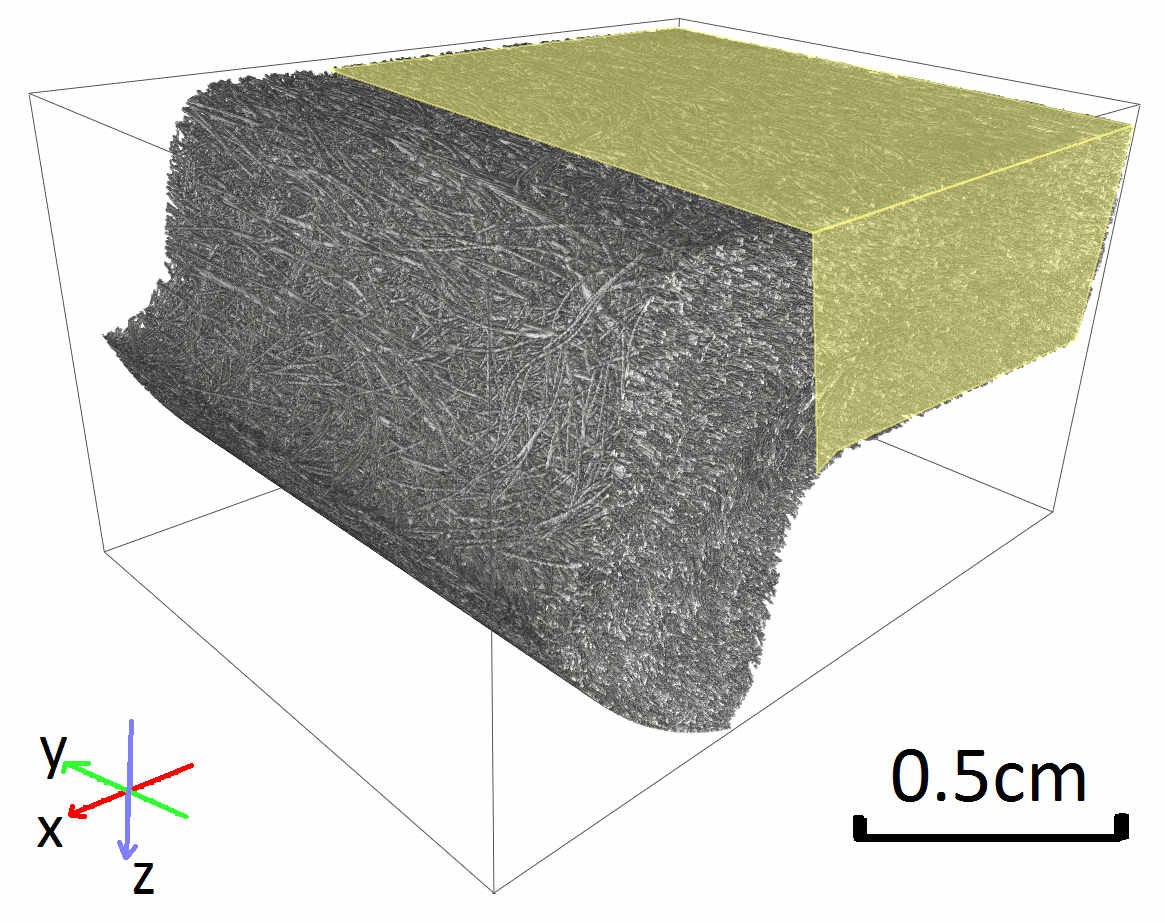}}\hfill
 \subfloat[A3.2uh, x-y-slice]{\includegraphics[width=\columnwidth]{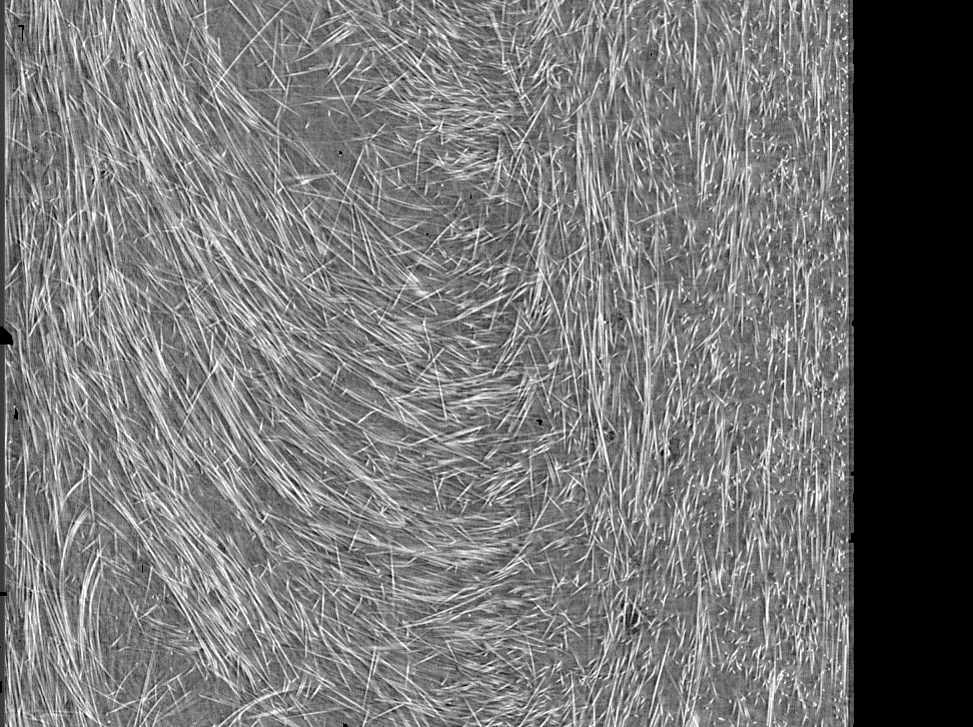}}
\caption{Visualizations of the reconstructed CT images of RoI A3.3 and A3.2uh, with pixel sizes of 3 and 10 \textmu m, respectively.}
\label{fig:vr-slices-ct-1}
\end{figure*}

All orientations are analyzed in 3D, in each voxel,
  see \cite{redenbach12:_beyon_imagin}. There, the Hessian matrix based method 
  is applied to data of FRP samples showing pixel-wise orientation results.  
  In \cite{weissenbock2014}, the authors discussed the existing problems 
  and challenges around the visualization of volumetric microstructures 
  by means of FRP. Size and
  complexity of the imaged regions necessitate a reduction of
  orientation information. Here, we concentrate
  on the orientation tensor diagonal element $a_{yy}$ as the
  $y$-direction is the dominating one. Therefore, results are averaged
  in cubic sub-volumes and often 2D slices or projections are chosen
  for better illustration.  Sub-volume sizes are nevertheless chosen
  with particular attention to formed layers, so called shell and core
  layers see \cite{hine2004, demonte2010b, demonte2010a}, in the components'
  microstructure. 

\subsection{RoI CT scans}
\label{sec:roi-ct-scans}
Altogether, four regions are scanned. Region A3 is scanned five times
with A3.1, A3.2m, A3.2h, A3.2uh, and A3.3 being ordered
w.r.t. ascending resolution. See Table~\ref{tab:regions} for
dimensions and voxel sizes.
\begin{figure*}[ht!]
 \subfloat[A4, volume rendering]{\includegraphics[width=\columnwidth]{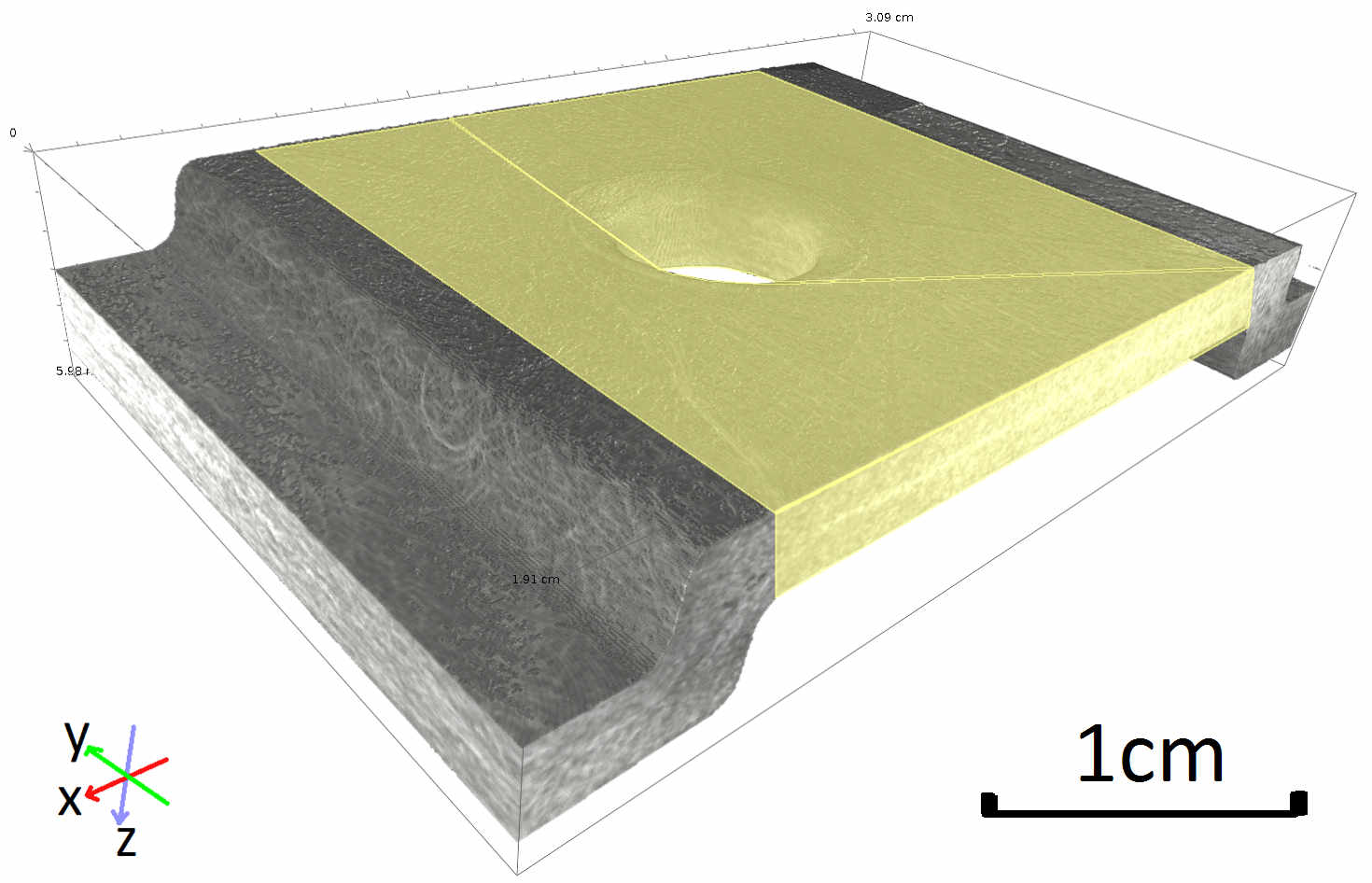}}\hfill
 \subfloat[A4, x-y-slice]{\includegraphics[width=\columnwidth]{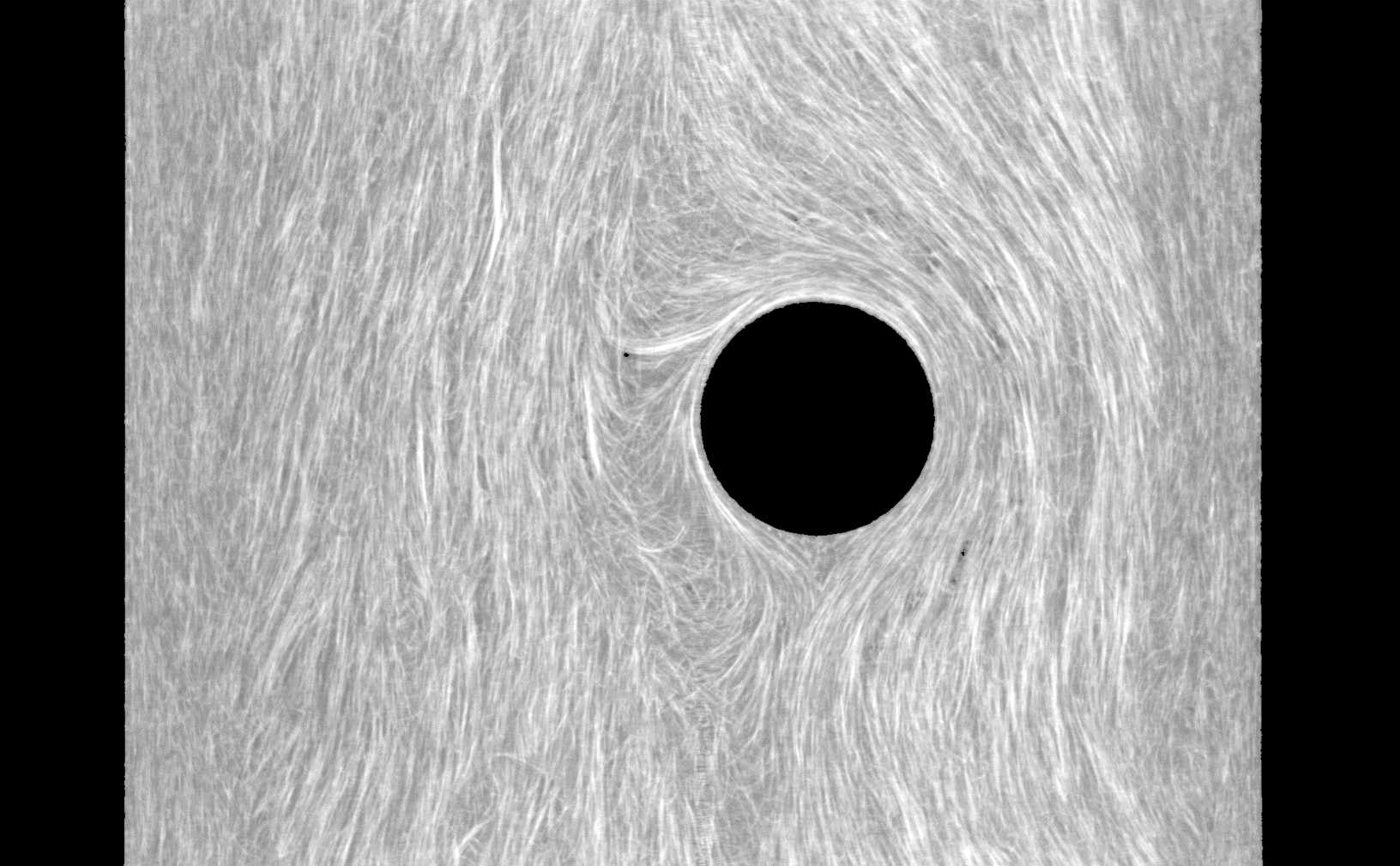}}\\
 \subfloat[A3.1, volume rendering]{\includegraphics[width=\columnwidth]{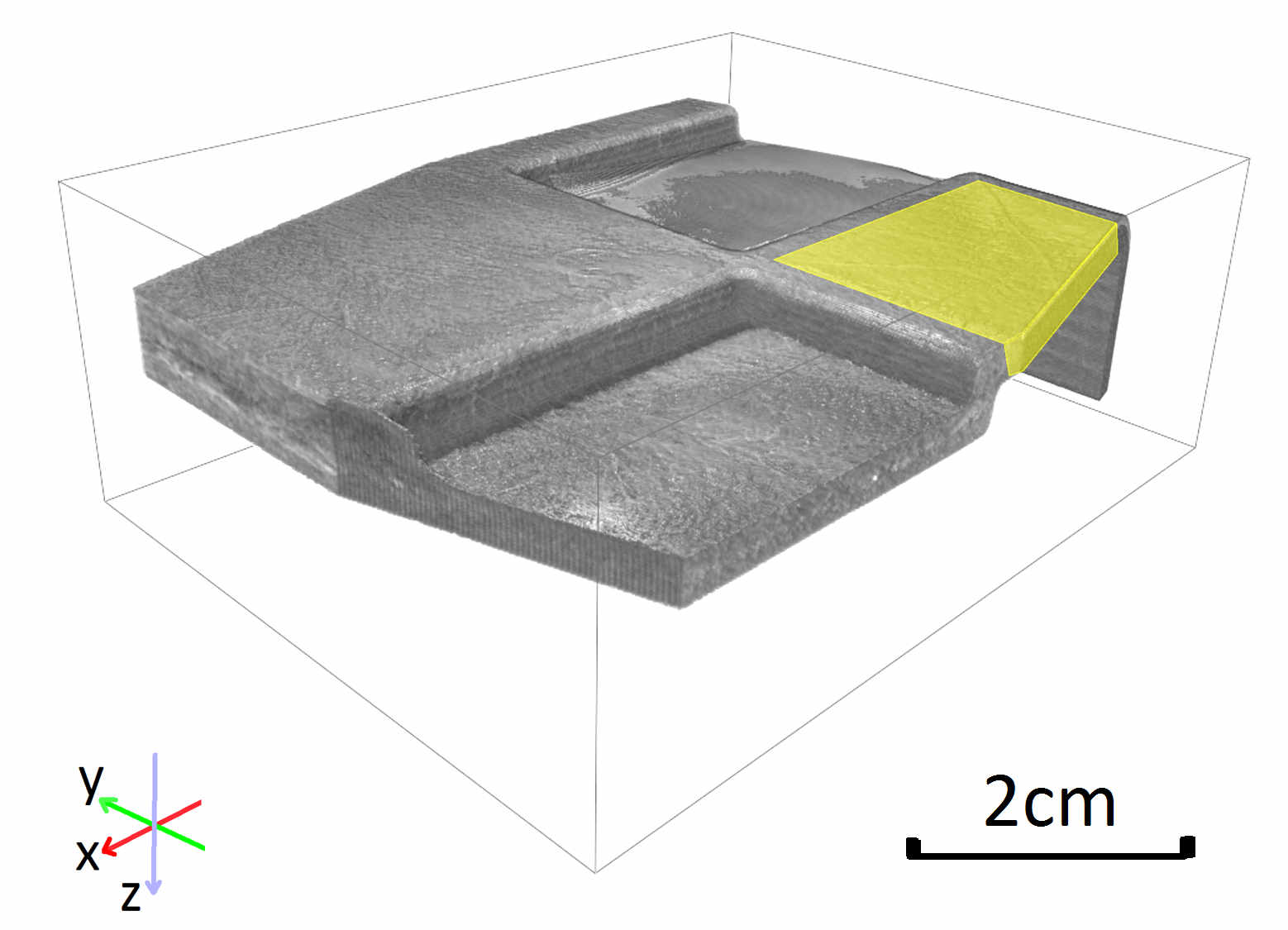}}\hfill
 \subfloat[A3.1, x-y-slice]{\includegraphics[width=\columnwidth]{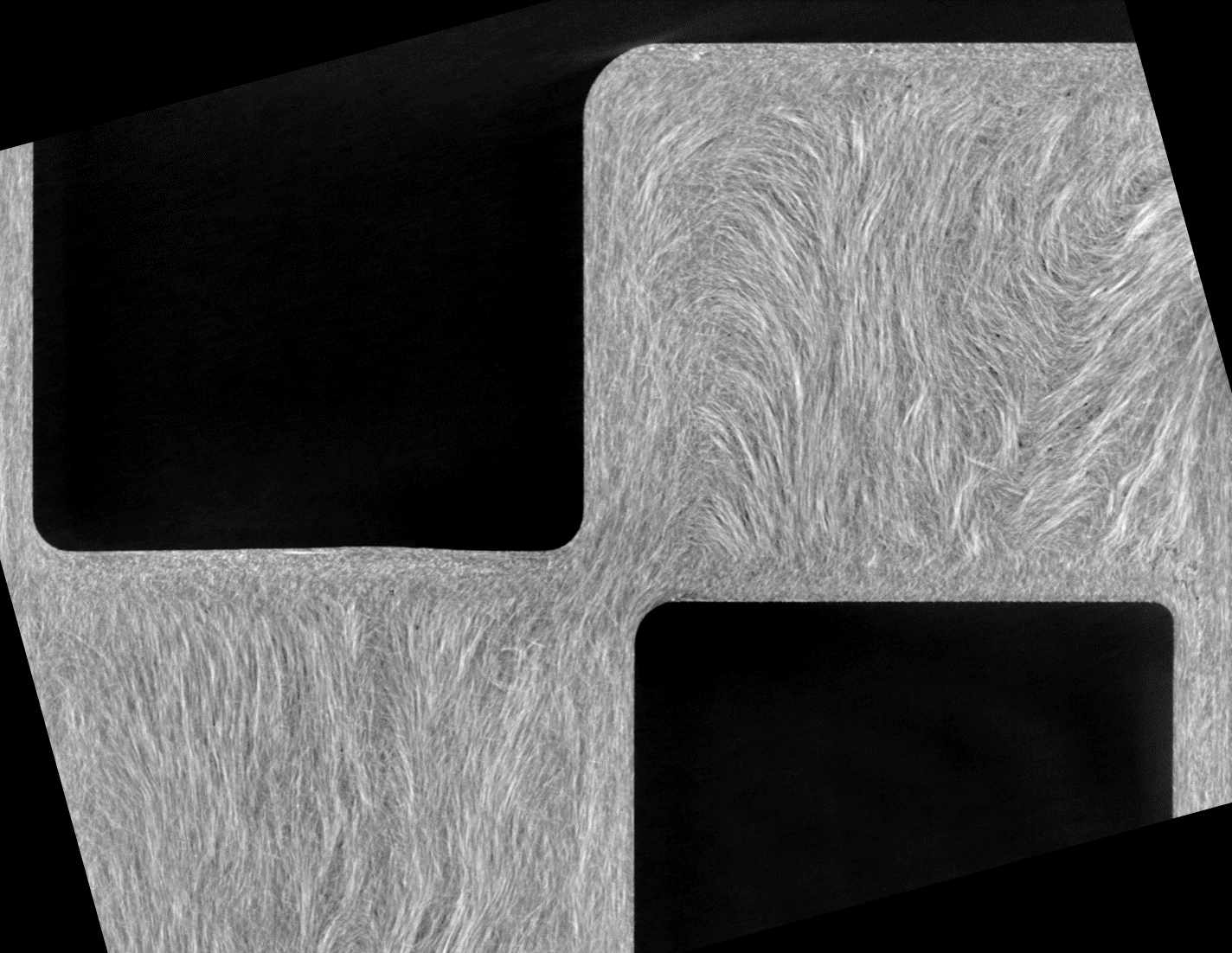}}
\caption{Visualizations of the reconstructed CT images of RoI A4 and A3.1, with pixel sizes of 17 and 44 \textmu m, respectively.}
\label{fig:vr-slices-ct-2}
\end{figure*}

Figures~\ref{fig:vr-slices-ct-1}-\ref{fig:vr-slices-ct-2} give a
visual impression of a representative selection of the analyzed
scans. Volume renderings and 2D slices of RoI A3.2uh, A4 and A3.1
reveal the differences in acquired data caused by varying resolutions
and the micro-structure in the imaged RoI respectively. The data sets
are cascaded in 
Figures~\ref{fig:vr-slices-ct-1}-\ref{fig:vr-slices-ct-2} starting at the
finest resolution and getting coarser towards the
bottom. Simultaneously this cascade emphasizes the differences in the
sample volume covered by the RoI, too. The 2D slices in 
Figures~\ref{fig:vr-slices-ct-1}-\ref{fig:vr-slices-ct-2} are taken from the
central layer of the plate.  All scans are subsequently
analyzed. However, local orientation analysis results are visualized
only for the selected regions marked yellow in 
Figures~\ref{fig:vr-slices-ct-1}-\ref{fig:vr-slices-ct-2}.
\subsection{Local fiber orientation tensors from 3D image analysis }
\label{sec:local-fiber-orient}
The computed orientation results are summarized in
Table~\ref{tab:CT-results}.  We derive the fiber component as
described in Section~\ref{sec:locFiberOrientation}.  The orientation
tensors are computed based on a tiling by cubes of size 218\,\textmu
m$\times$218\,\textmu m$\times$218\,\textmu m. Clearly the y-direction
is preferred (long axis of the carrier) for regions A3.1, A3.2, A3.3
and A5. In RoI A4, the y-direction is less dominant due to the
reorientation around the hole in this region.  RoI A2 is a special
case as the component's shape differs significantly from the remaining
plate-like shape in this spur region. A2 features $a_{xx}$ as the
highest component. This finding is not surprising due to x being the
longitudinal direction of the spur, see Figure
\ref{fig:ROI_Ford_Bauteil}.
\begin{table*}[ht!]
\centering
 \begin{tabular}{|c|c|c|c|c|}
  \hline
  RoI & mean fiber & anisotropy & orientation tensor & voxel size \\
	& direction & index & $a_{xx}$, $a_{yy}$, $a_{zz}$ & $[$\textmu m$]$ \\
  \hline\hline 
       A2 & not applicable & $0.55$ & $0.52,\ 0.23,\ 0.23$ & $45$\\ \hline
       A3.1 & not applicable & $0.55$ & $0.25,\ 0.51,\ 0.23$ & $44$ \\ 
       A3.2m & $\left(-0.07,\ \phantom{-}0.99,\ -0.00\right)^T$ & $0.61$ & $0.23,\ 0.54,\ 0.21$ & $45$\\
       A3.2h & $\left(-0.08,\ \phantom{-}0.99,\ -0.02\right)^T$ & $0.60$ & $0.22,\ 0.54,\ 0.22$ & $21$\\
       A3.2uh & $\left(-0.06,\ \phantom{-}0.99,\ -0.05\right)^T$ & $0.65$ & $0.21,\ 0.58,\ 0.21$ & $10$\\ \hline    
       A3.3 & $\left(-0.05,\  -0.99,\  -0.02\right)^T$\parbox[b][14pt][b]{0pt}{} & $0.78$ & $0.18,\ 0.66,\ 0.17$ & $3$ \\  \hline  
       A3.3.1 & $\left(-0.11,\ -0.99,\ \phantom{-}0.03\right)^T$\parbox[b][14pt][b]{0pt}{} & $0.75$ & $0.18,\ 0.63,\ 0.18$ & $3$\\ 
       A3.3.2 & $\left(-0.03,\ -0.99,\ -0.03\right)^T$ & $0.78$ & $0.16,\ 0.67,\ 0.16$ & $3$\\ 
       A3.3.3 & $\left(-0.08,\ -0.99,\ -0.04\right)^T$ & $0.76$ & $0.18,\ 0.64,\ 0.16$ & $3$\\ 
       A3.3.4 & $\left(-0.05,\ -0.99,\ -0.05\right)^T$ & $0.78$ & $0.17,\ 0.65,\ 0.16$ & $3$\\ 
       A3.3.5 & $\left(-0.04,\ -0.99,\ -0.04\right)^T$ & $0.77$ & $0.18,\ 0.63,\ 0.17$ & $3$\\ 
       A3.3.6 & $\left(\phantom{-}0.01,\ \phantom{-}0.99,\ \phantom{-}0.04\right)^T$ & $0.77$ & $0.19,\ 0.63,\ 0.17$ & $3$\\ \hline 
       A4 & not applicable & $0.50$ & $0.26,\ 0.48,\ 0.24$ & $17$\\ \hline
       A5 & not applicable & $0.57$ & $0.22,\ 0.51,\ 0.25$ & $44$\\ \hline
 \end{tabular}
 \caption{Orientation analysis results for the
   entire RoI, as specified in Table
   \ref{tab:regions}. Note that the mean fiber direction is indicative only if the anisotropy index exceeds
   0.6. The respective vectors are therefore not reported if the anisotropy is below this bound.}
\label{tab:CT-results}
\end{table*} 

\begin{figure}[!ht]
 \includegraphics[width=\columnwidth, trim=0mm 0mm 0mm 20mm]{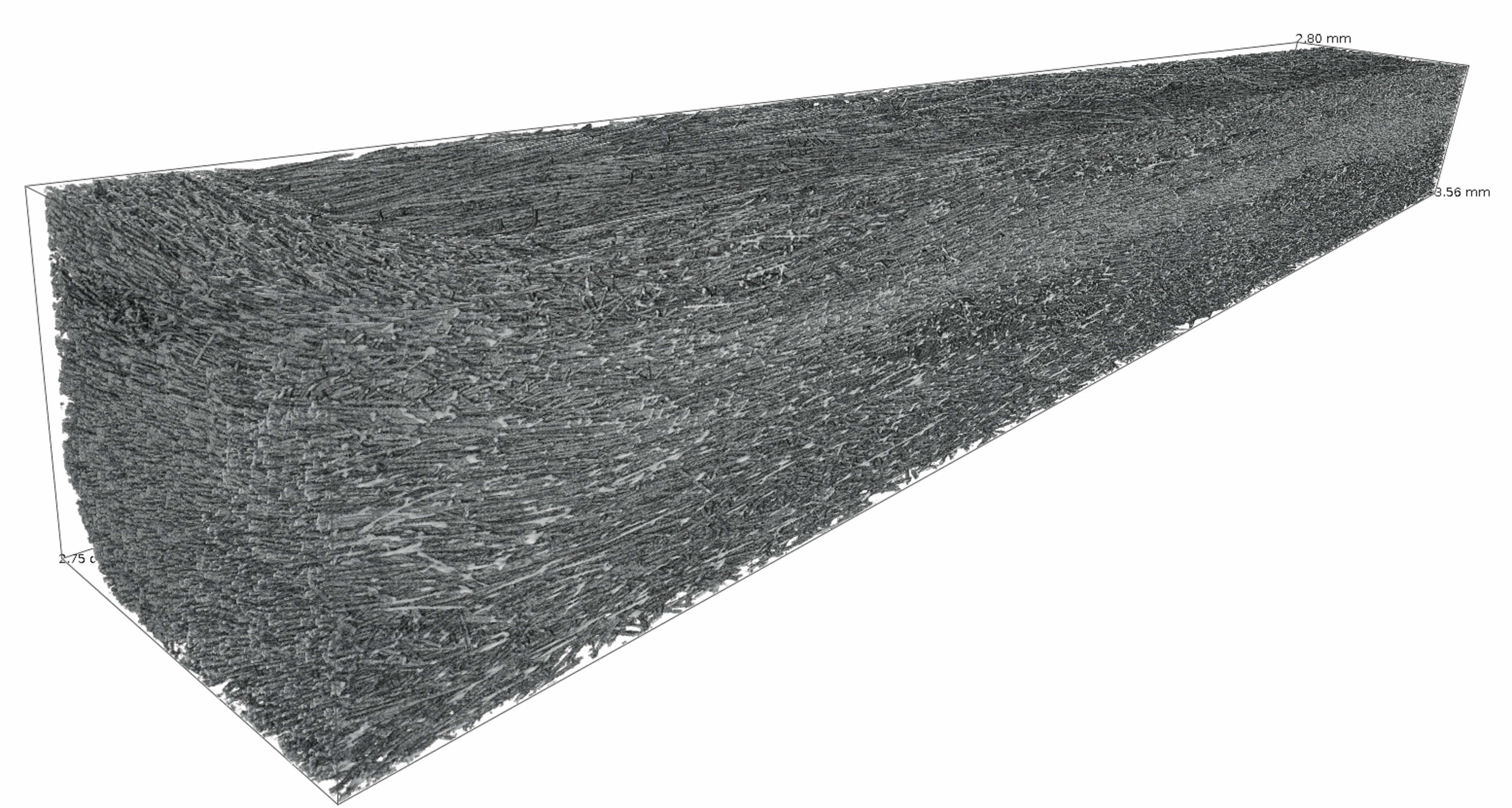}
 \caption{A3.3, cropped to the plate region, resulting in a
   $1\,200\times 8\,832\times 1\,000$ voxel sub-volume. The area
   closest to the reader is close to an edge where fibers are forced
   to bend.}
\label{fig:roiA3p3_analyzedMask}
\end{figure}

\begin{figure}[!ht]
 \includegraphics[width=\columnwidth]{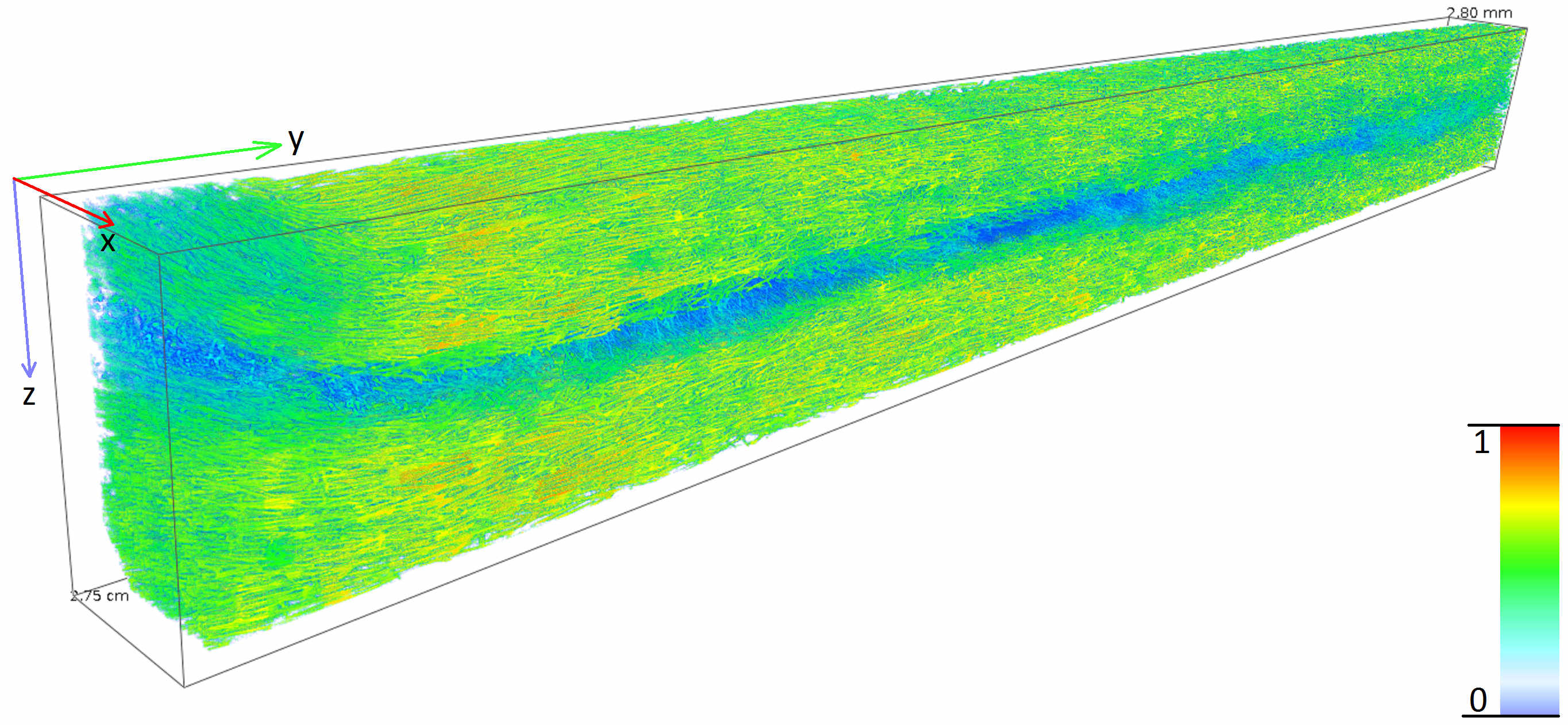}
 \caption{Clipped visualization of the evaluated area. The orientation tensor component $a_{yy}$ is shown in the analyzed sub-volumes of edge length 65 pixels (about 200 \textmu m) leading to a grid of 17$\times$135$\times$13 cubes. Averaged over the whole analyzed volume, the orientation tensor diagonal components are $a_{xx}=0.20,\ a_{yy}=0.63,\ a_{zz}=0.15$. The anisotropy index is $0.75$ and the mean fiber direction $\left(-0.00,\ -0.99,\ -0.01\right)^T$.}
\label{fig:roiA3p3_rendering_ayy}
\end{figure}

\begin{figure}[!ht]
  \includegraphics[width=\columnwidth]{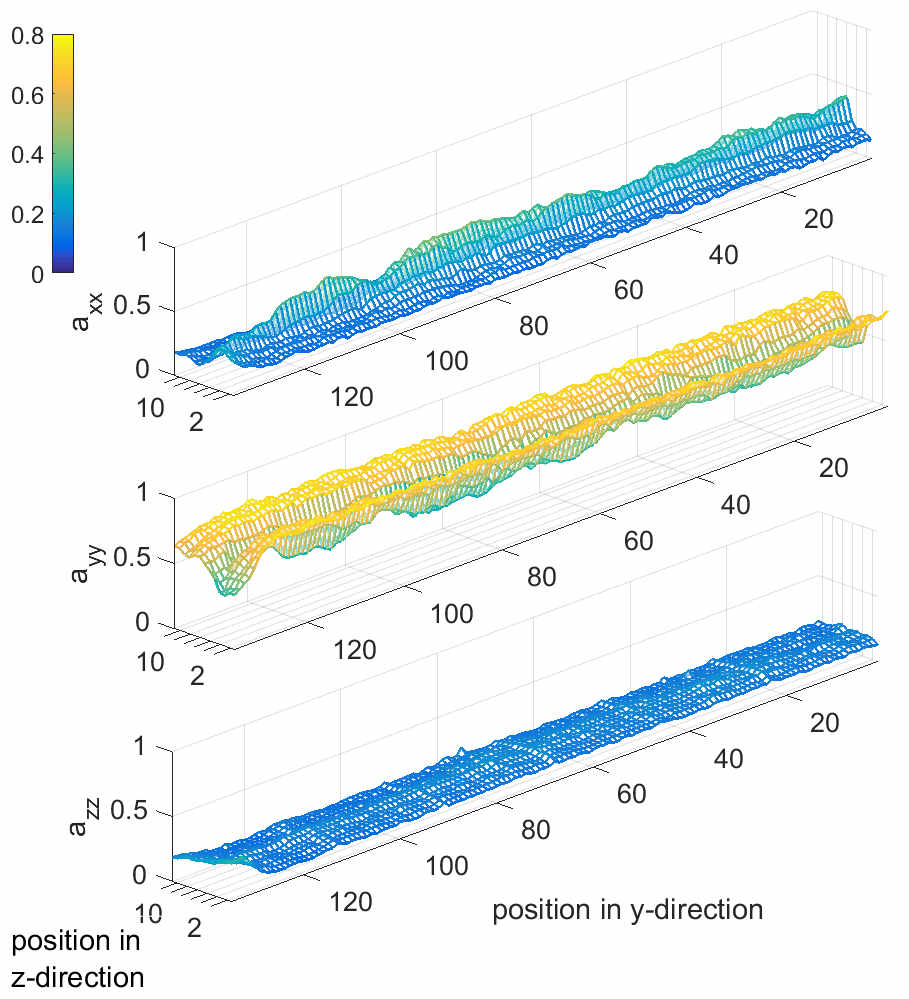}
 \caption{Orientation tensor diagonal components for the RoI scan
   A3.3. The mesh is based on the analysis in 135$\times$13
   sub-volumes, where the results are averaged over 17 sub-volumes
   along the x-axis.}
\label{fig:roiA3p3_analyzed_sffd65}
\end{figure}

In the following Figure~\ref{fig:roi_analyzed_sffd5}, we show the
local orientation results for the regions marked yellow in
Figures~\ref{fig:vr-slices-ct-1}-\ref{fig:vr-slices-ct-2}. The Figure
shows slices from the upper (left) and central (right) layers of the
plate like sub-regions.  The color map visualizes the orientation
tensor diagonal element $a_{yy}$ in flow direction y. For all
resolutions, as expected, fiber orientations cluster around the flow
direction in the upper layer and deviate stronger from this preferred
direction in the central layer. Note that the orientation results in
Table~\ref{tab:CT-results} are averaged over the entire regions. Thus,
fibers being reoriented along the edges due to shaping are taken into
account, too.  This might decrease the anisotropy index as well as the
dominating diagonal orientation tensor element. The yellow marked
regions are tightly limited to the areas in plane and thus avoid the
reorientation of fibers due to shaping procedures. Orientation results
for these regions are visualized in
Figure~\ref{fig:roi_analyzed_sffd5} for the coarse resolution scans
and the sequence of
Figures~\ref{fig:roiA3p3_analyzedMask}-\ref{fig:roiA3p3_analyzed_sffd65}
for the finest resolution scan.
\begin{figure*}[!ht]
 \subfloat[A3.2uh, upper layer]{\includegraphics[width=\columnwidth]{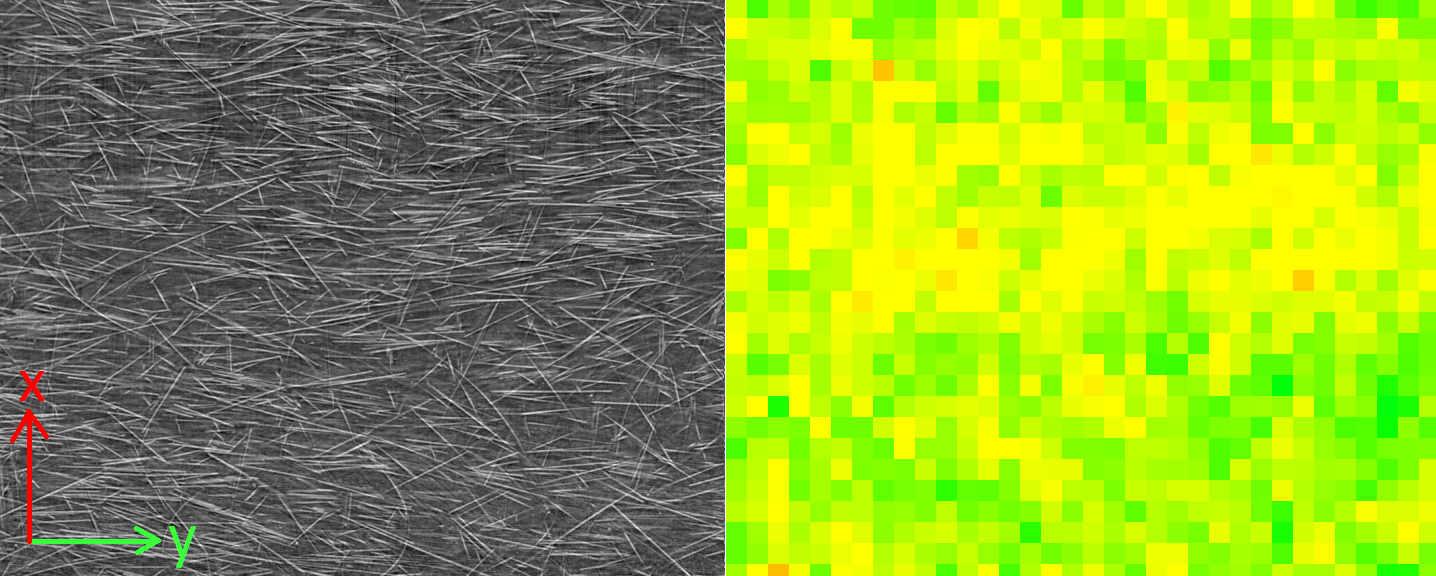}}\hfill
 \subfloat[A3.2uh, center layer]{\includegraphics[width=\columnwidth]{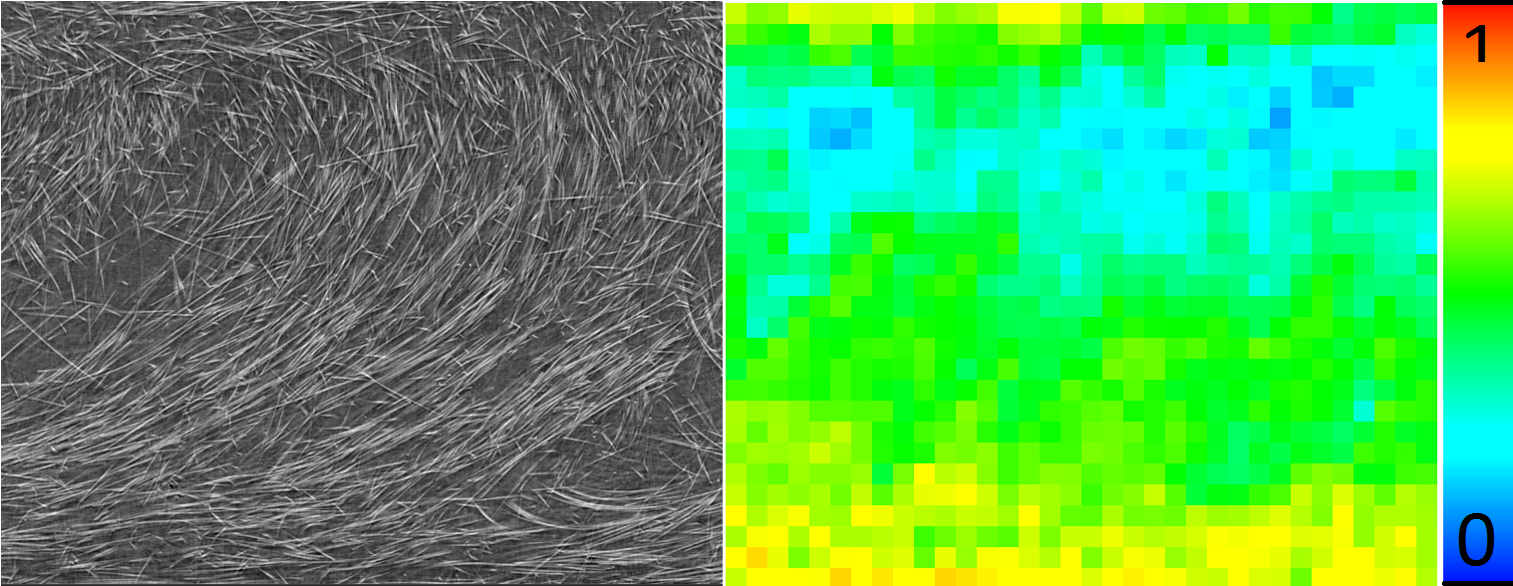}} \\
 \subfloat[A4, upper layer]{\includegraphics[width=\columnwidth]{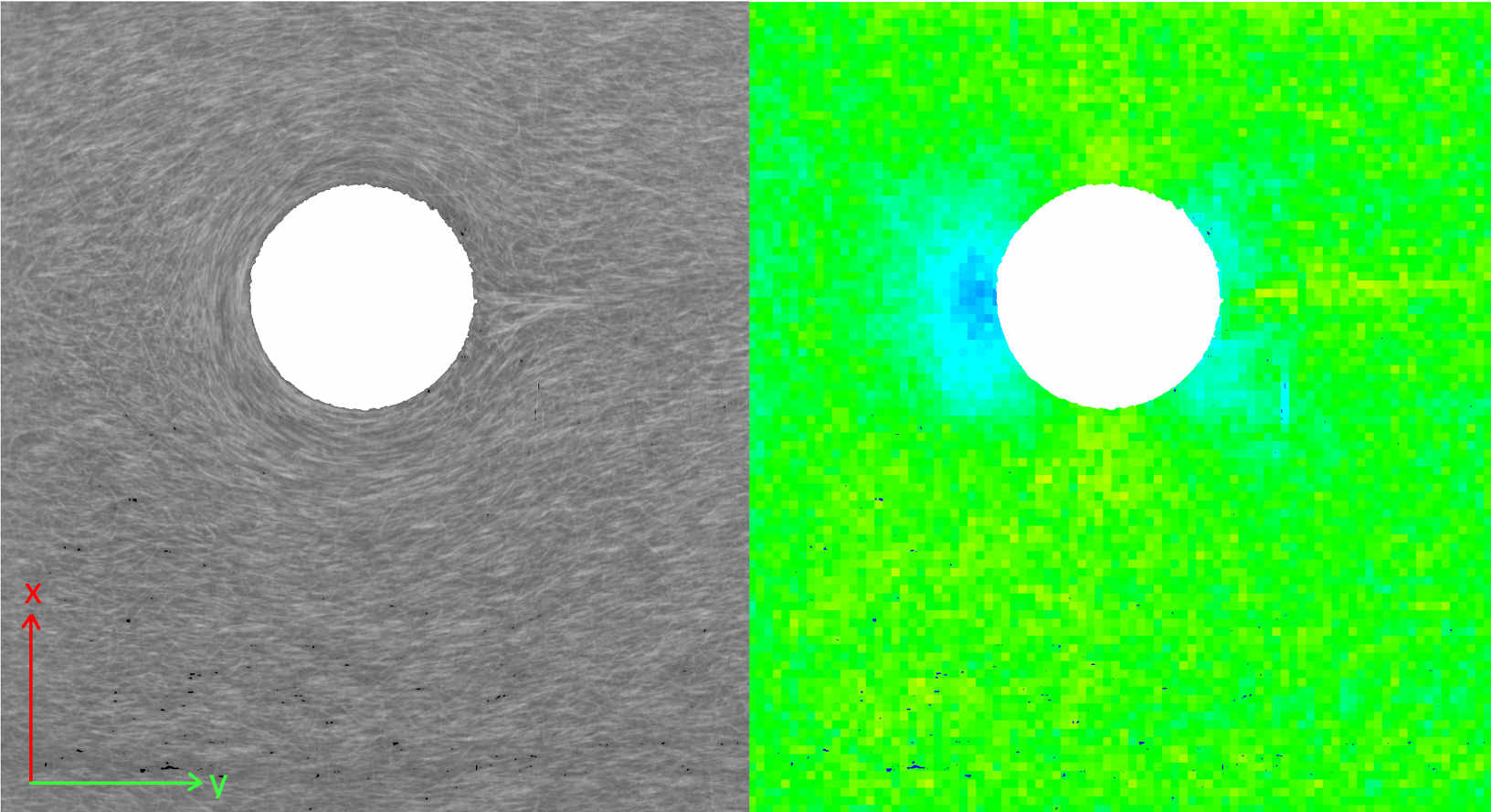}}\hfill
 \subfloat[A4, center layer]{\includegraphics[width=\columnwidth]{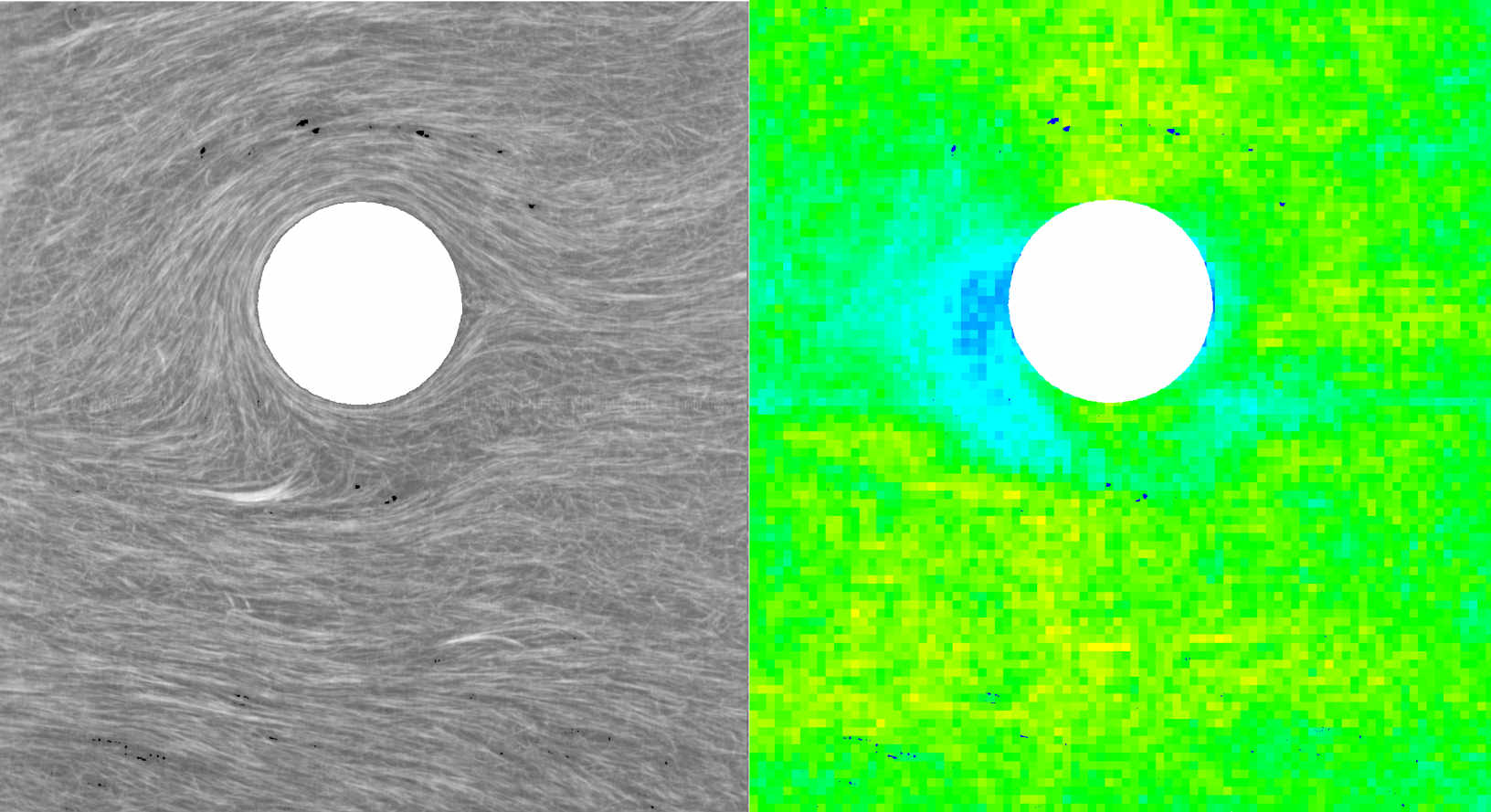}} \\
 \subfloat[sub-sample of A3.1, upper layer]{\includegraphics[width=\columnwidth]{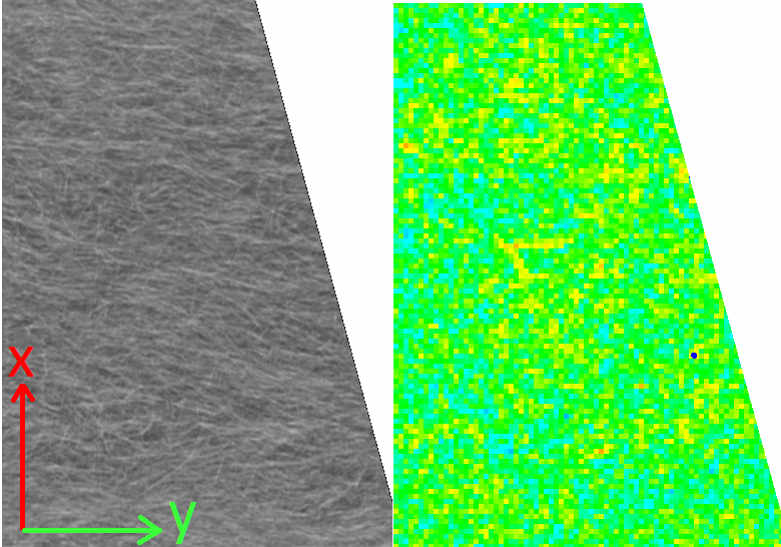}}\hfill
 \subfloat[sub-sample of A3.1, center layer]{\includegraphics[width=\columnwidth]{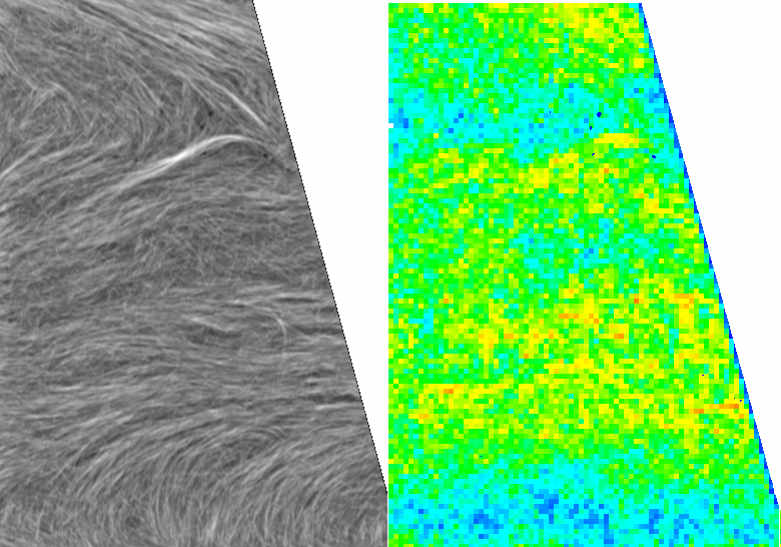}} 
 \caption{Orientation estimation results for three regions marked yellow in
   Figures~\ref{fig:vr-slices-ct-1}-\ref{fig:vr-slices-ct-2}. 2D
   slices from the original 3D images and from the image holding the
   corresponding computed local tensor component $a_{yy}$ are shown
   next to each other. The orientation tensors are computed in cubic
   sub-volumes of (218\,\textmu m)$^3$ for all data sets.}
\label{fig:roi_analyzed_sffd5}
\end{figure*}  

Figures~\ref{fig:roiA3p3_analyzedMask} and
\ref{fig:roiA3p3_rendering_ayy} show the ‘edge’ in region A3 which was
scanned at the highest resolution (3\,{\textmu}m/voxel). The
orientation tensor is averaged in boxes of edge length
200\,{\textmu}m. The expected central layer deviating from the
dominating y-orientation \cite{musiko2019} is clearly visible
in the rendering. Figure~\ref{fig:roiA3p3_analyzed_sffd65} shows the
orientation tensor diagonal elements, averaged along the x-axis of the
shown volume. These graphs reveal that the central layer has a
pronounced x-orientation ($a_{xx}$ rising from 0.2 to approximately
0.5 along the entire strap) at the expense of y-alignment (dropping
from 0.6 to 0.3) while the $a_{zz}\approx 0.2$ tensor component is
constant over the entire thickness
(Figure~\ref{fig:roiA3p3_analyzed_sffd65}(c)).  Moreover,
Figure~\ref{fig:roiA3p3_analyzedMask} indicates a slight reorientation
of fibers at one ending of the elongated volume. This visual
impression is backed by Figure~\ref{fig:roiA3p3_analyzed_sffd65}.
Clearly, fibers are oriented mostly in plane in the first 120
sub-volumes along the y-axis in contrast to the last 15 sub-volumes,
where higher $a_{zz}$ values are observed. This is exactly where the
fibers are reoriented due to shaping. 

\subsection{Comparison of local fiber orientation tensors derived from
scans at varying resolutions}
\label{sec:comp-resolution}
\begin{figure}[!ht]
\includegraphics[width=\columnwidth]{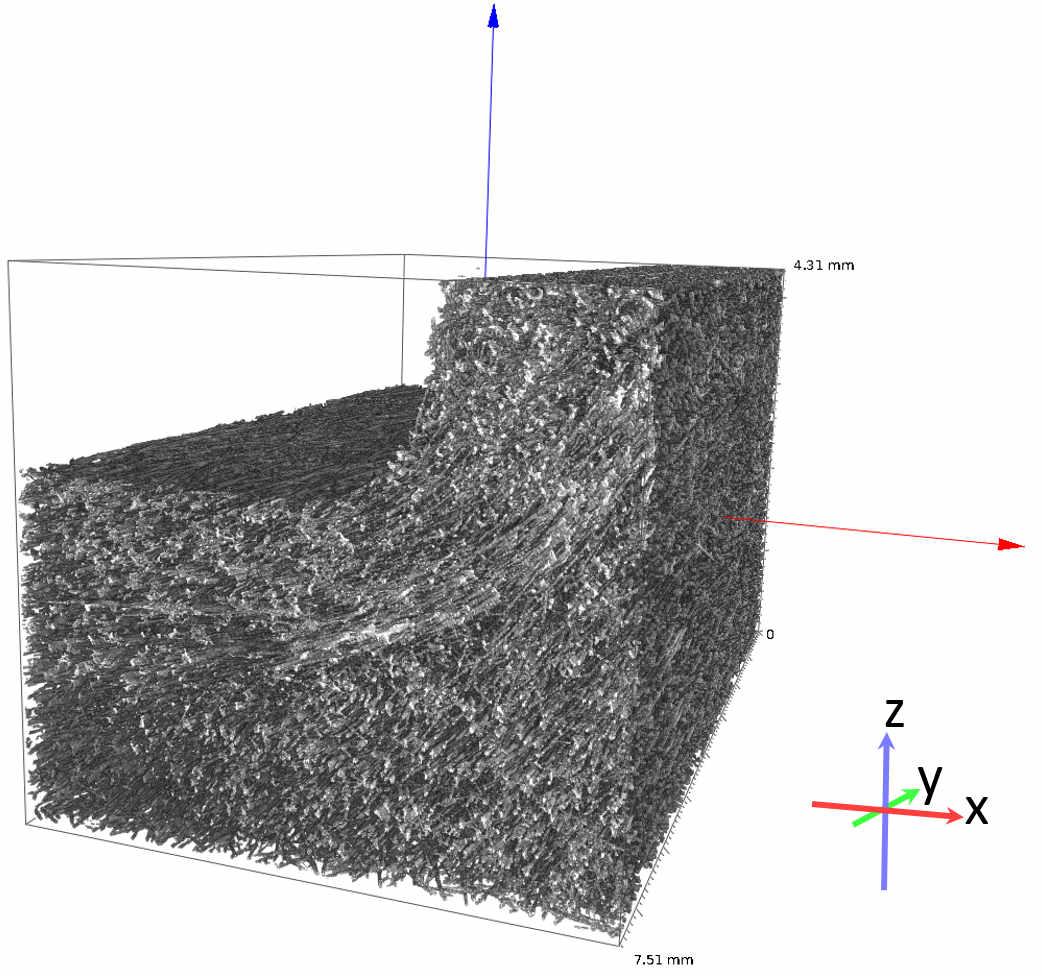}
\caption{Volume rendering of the sub-volume used for the comparison in
  Figure~\ref{fig:A3-local-comparison-plots}.}
\label{fig:A3-local-comparison-vr}
\end{figure}

\begin{figure*}[!hbt]
\includegraphics[width=\textwidth]{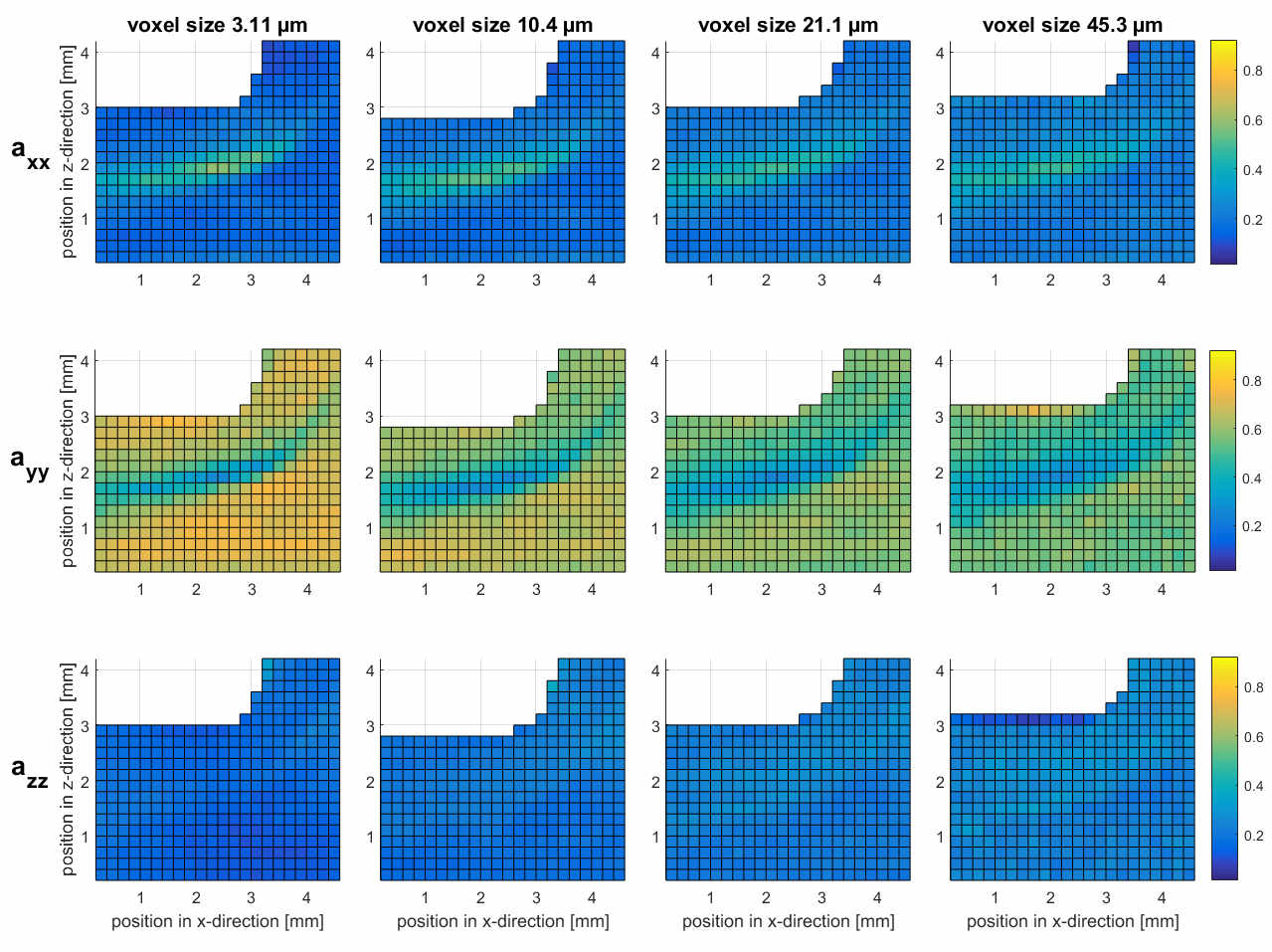}
\caption{Comparison of local results for varying voxel sizes:
  Orientation tensor diagonal components for the RoI scan, color
  coded. A volume rendering of the corresponding sub-volume is shown
  in Figure~\ref{fig:A3-local-comparison-vr}.}
\label{fig:A3-local-comparison-plots}
\end{figure*}

In order to compare the local orientations derived from scans at
varying lateral resolutions quantitatively, we chose a sub-volume of
A3 that is covered by several scans.  More precisely, the chosen
volume lies in the intersection of the RoIs A3.3, A3.2uh, A3.2h, and
A3.2m, scanned with voxel sizes 3-45\,\textmu m. See
Figure~\ref{fig:A3-local-comparison-vr} for a volume rendering. We
averaged the orientation results in sub-volumes of edge-length
200\,\textmu m and subsequently along the y-axis in order to preserve
the characteristic differences between shell and core
layers. Figure~\ref{fig:A3-local-comparison-plots} shows the
remarkable consistency of the orientation results even for the
coarsest resolution at 45.3 \textmu m.  Nevertheless, the quality of
orientation results drops for the coarser resolutions. This becomes
obvious by the lower color contrasts between shell and core layers
from left to right. Moreover, the component $a_{zz}$ orthogonal to the
plate varies the most when resolved at 45.3 \textmu{m}. These two
observations clearly show a bias towards isotropy in the analysis
results.

\subsection{Comparison to local fiber orientation tensors from
  injection molding simulation}
\label{sec:comp-local-fiber}
\begin{figure*}[!ht]
\begin{center}
 \includegraphics[width=0.75\textwidth]{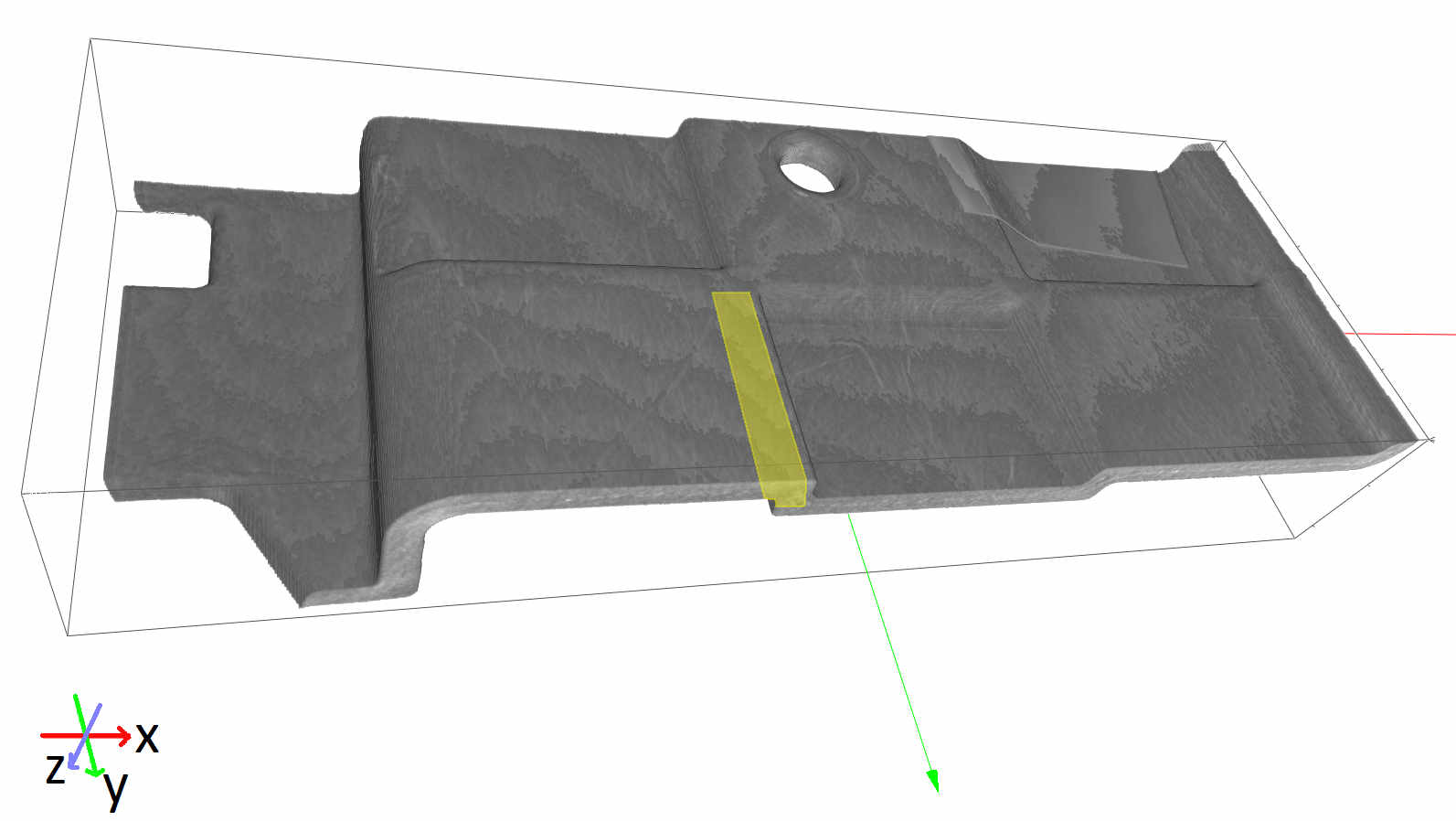}    
\end{center}
 \caption{Visualization of a coarse CT scan around RoI A3.3 (marked
   yellow) that has been used for injection simulation.}
\label{fig:coarseCTscan}
\end{figure*}

\begin{figure*}[!hbt]
\begin{center}
 \includegraphics[width=0.8\textwidth]{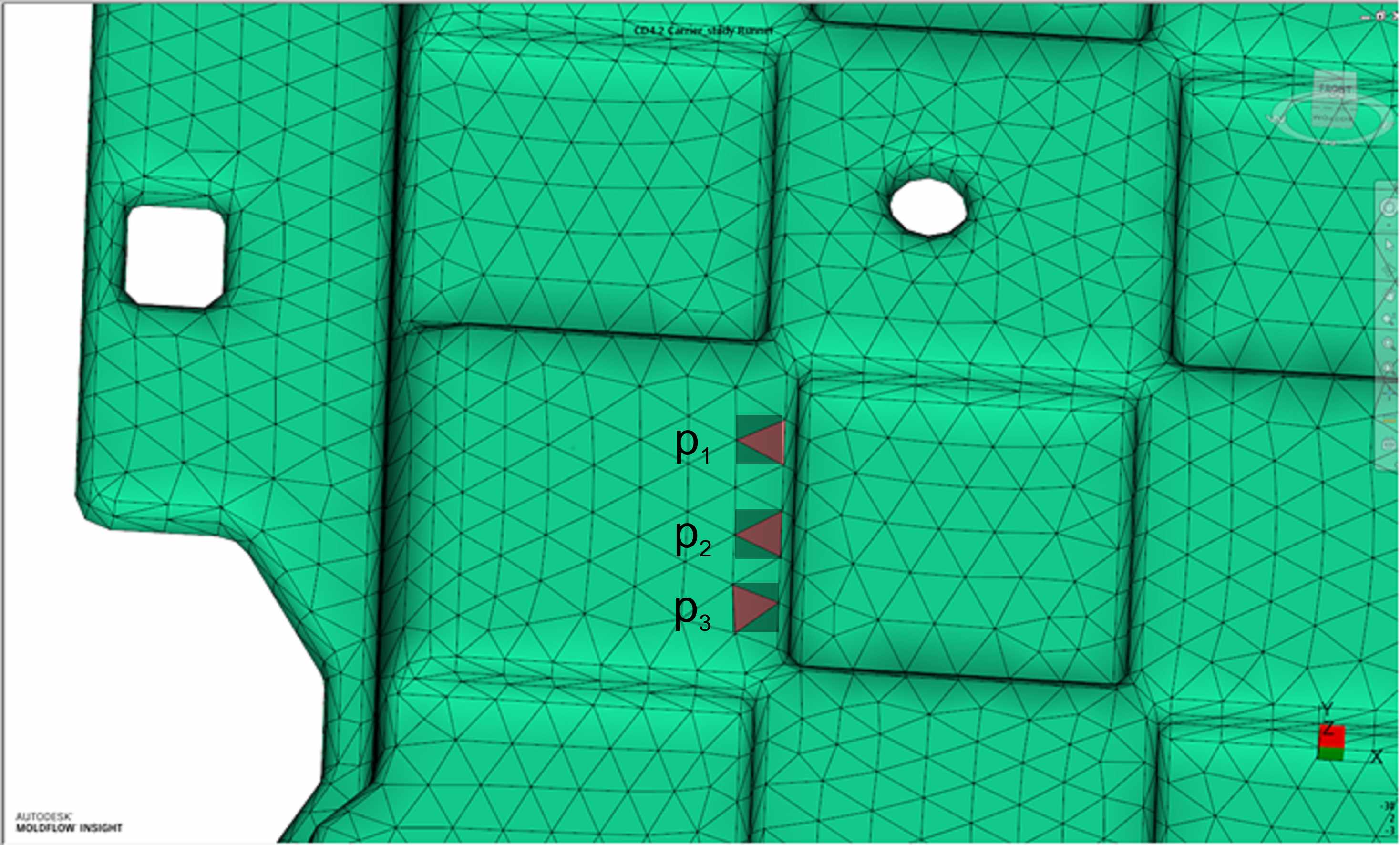}
 \end{center}
 \caption{Mesh of the area around A3.3 that has been used for
   injection simulation in the software Autodesk Moldflow Insight
   2017. p$_1$, p$_2$ and p$_3$ mark the positions where the
   simulation results have been extracted. The simulated orientation tensor is averaged in the marked triangle, whereas the analysis based on the
   CT data averages in cubes.}
\label{fig:moldflow_mesh}
\end{figure*}

For three sub-regions of A3, we compare the 2nd order orientation
tensors calculated in the previous section with the values obtained by
Moldflow\textsuperscript{\textregistered} simulations. The region is
highlighted in yellow in Figure~\ref{fig:coarseCTscan}.
The compared positions p$_1$ to p$_3$ are
marked by triangles and rectangles in Figure~\ref{fig:moldflow_mesh} for
simulation and \textmu CT data respectively.
The tetrahedral Moldflow mesh can be seen in
Figure~\ref{fig:moldflow_mesh} along with three tetrahedra for which
we calculated orientation tensors for comparison with the results from
{\textmu}CT.

\begin{figure*}[!hbt]
 \includegraphics[width=1.0\textwidth]{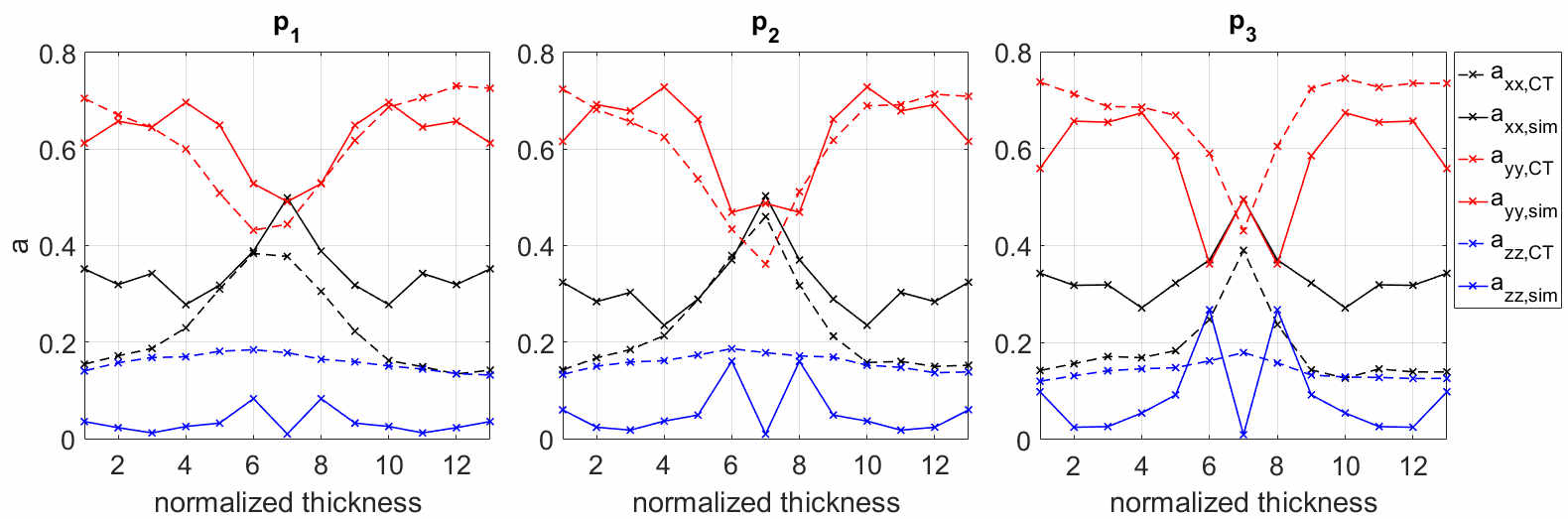}
 \caption{Comparison of simulated and CT image data based orientation
   tensors for region A3.3.}
\label{fig:comparison_A3p3}
\end{figure*}

The used mesh size is app. 2mm (element edge length in-plane). 12
layers are used over the part thickness (out-of-plane) for the
calculations. The entire carrier has around 170\,000 elements using
Dual Domain mesh.  For the orientation calculation, the Moldflow
Rotational Diffusion model is used, see
e. g. \cite{tseng-thermoplastics-2018}. Its parameters -- fiber
interaction coefficient Ci and coefficients of asymmetry $D1,\ D2$ and
$D3$ -- are set as “automatic calculations”. That is, the default
values $D1=1.0, D2=0.8, D3=0.15$ are used. The \textmu CT data is
processed in a way to fit the grid of simulated data. The plate
thickness is subdivided in 12 cuboidal sub-volumes. The orientation
results are obtained by averaging along x- and y-axis (in plane).

Figure~\ref{fig:comparison_A3p3} shows the three diagonal tensor
elements $a_{xx}$, $a_{yy}$, and $a_{zz}$ from both the simulation and
the measurement. The components $a_{xx}$ and $a_{yy}$ agree
qualitatively well with $a_{xx}$ peaks in the central layer and
decreasing to the outside and $a_{yy}$ showing the opposite behavior
for both methods. The $a_{zz}$ component however is more pronounced in
some measurement points in the simulation while it is almost
negligible throughout in the measurement with a slight exception
around the center of p$_3$. In particular the $a_{yy}$ and $a_{zz}$
components feature kinks in the central layer that are neither
explained by standard flow dynamics nor backed by the {\textmu}CT
measurements.

\section{Conclusion}
Based on an off-the-shelf part from the automotive industry, we showed
that RoI CT is readily applicable to large FRP components provided
that the CT scanner has the required motor axes and degrees of
freedom. We scanned an injection molded automotive glass FRP component
that is 0.9\,m long, 0.35\,m wide, and has a wall thickness of 2\,mm. 

It is commonly assumed that a single fiber has to be sampled by at
least 2-4 voxels per diameter, see the corresponding
discussion in the Introduction. We clearly proved this assumption to be
false.  Sampling the fiber diameter by less than a voxel still allows
for 3D orientation analysis in each voxel predominantly representing
the fiber phase. The analysis method of choice gains local orientation
information in each voxel from the gray value curvature captured by
the Hesse matrix.  The results are stabilized by averaging the 2nd
order orientation tensor over small sub-volumes. By this method we
achieved good and consistent results for voxel sizes between
45\,{\textmu}m and 3\,{\textmu}m for the same material.  Measuring the
local orientation of individual fibers would require a much higher
spatial sampling and therefore impose a much smaller FoV.

The 3D image analysis reveals strong anisotropy in the local fiber
orientation. For all RoI except the spur shaped A2, the injection direction
y is as expected the preferred one, see Table~\ref{tab:CT-results}.
Local effects like the typical thickness-dependent changes in the mean
fiber orientation caused by flow turbulences in the central layer are
captured, regardless the image quality, see
Figures~\ref{fig:roi_analyzed_sffd5} and
\ref{fig:A3-local-comparison-plots}. 
Caution is however advised when
it comes to a quantitative comparison as blurred structural
information induces a bias of the orientation analysis results towards
isotropy, see the right
column of Figure~\ref{fig:A3-local-comparison-plots}. Comparing the 
analysis results for A3.1 and A3.2, see Table \ref{tab:CT-results}, shows that
nominal resolution is however not the decisive parameter here.

For the best resolved RoI {\textmu}CT scan A3.3, the local fiber
orientations observed in the RoI CT data can be compared to those
derived by injection molding simulation
(Figure~\ref{fig:comparison_A3p3}). Although deviating quantitatively,
the results do agree qualitatively and particularly well for the
dominating tensor component. This proves clearly that the measured
orientation results can be used to validate simulation results. Thus
RoI {\textmu}CT combined with 3D image analysis as applied here,
enables truly non-destructive 3D micro-structure characterization for
large FRP components.

To summarize, we proved RoI {\textmu}CT to be a potential standard
tool for local fiber orientation analysis in glass fiber-reinforced
automotive parts. Also, we applied successfully the 
proposed RoI scanning technique in combination with the orientation 
analysis method to carbon FRP in \cite{zabler2019}. 
Our results are very encouraging and suggest that
the method is applicable to both short and long fiber reinforced
composites. Unlike methods which try to find single individual fibers
and which therefore require a very high resolution and consequently
cover a very small measurement volume, our method allows the extension
of the latter to some 2\,000 times the fiber diameter, hence 3\,cm for
15\,{\textmu}m thick glass fibers.

\subsection*{Acknowledgement}
This work has been supported by the Fraunhofer Society under project
MEF 3D Volant. DD has been supported by the German Federal Ministry of
Education and Research (BMBF) under grant 05M13RCA. 

\bibliographystyle{abbrv} 
\bibliography{litbank,ct-refs,ct-frp}

\end{document}